\documentclass{aa}  

\usepackage{graphicx}
\usepackage{txfonts}

\usepackage{color}
\usepackage{array}
\usepackage{caption}
\usepackage{subcaption}
\usepackage{amsmath,amssymb,amsfonts}
\usepackage{gensymb, footnote}
\usepackage{tikz}
\usepackage{sidecap}
\usepackage{changepage}
\usepackage{tabu}
\usepackage[toc,page]{appendix}
\usepackage{multirow}
\usepackage[hidelinks]{hyperref}

\usepackage{tablefootnote}
\usepackage{threeparttable}

\newcommand{\Teff}{T_{\text{eff}}}
\newcommand{\Teffo}{T_{\text{eff,\text{B}}}}
\newcommand{\Tefft}{T_{\text{eff,\text{C}}}}
\newcommand{\Mini}{M_{\text{initial}}}
\newcommand{\Modot}{M_{\odot}}
\newcommand{\MSun}{M_{\odot}}

\newcommand{\logL}{\log(L/L_{\odot})}

\newcommand{\kms}{km\,s$^{-1}$}

\newcommand{\Tabref}{Table\,\ref}
\newcommand{\Figref}{Fig.\,\ref}
\newcommand{\Secref}{Sect.\,\ref}
\newcommand{\Eqref}{Eq.\,\ref}

\newcommand{\nextline}{\\\indent}

\newcommand{\BAT}{BAT99\,126}

\begin{document}

   \title{BAT99\,126: A multiple Wolf-Rayet system in the Large Magellanic Cloud with a massive near-contact binary\thanks{Based on observations made with ESO telescopes at the La Silla Paranal Observatory under programme ID  0102.D-0050(A)}} \titlerunning{BAT99\,129: a multiple WR system}

   \author{S. Janssens
          \and
          T. Shenar
          \and 
          L. Mahy
          \and 
          P. Marchant
          \and
          H. Sana
          \and 
          J. Bodensteiner 
          }

   \institute{Institute of Astronomy, KU Leuven, Celestijnenlaan 200D, 3001 Leuven, Belgium\\
              \email{soetkin.janssens@kuleuven.be}}

   \date{Received 1 September 2020; accepted 14 November 2020}

 
  \abstract
   {\BAT\space is a multiple system in the Large Magellanic Cloud containing a Wolf-Rayet (WR) star, which has a reported spectroscopic (orbital) period of 25.5\,days and a photometric (orbital) period of 1.55\,days, and hence is potentially one of the shortest WR binaries known to date. Such short-period binary systems that contain a WR star in low-metallicity environments are prime candidate progenitors of black-hole (BH) mergers.
   }
   {By thoroughly analysing the spectroscopic and photometric data, we aim to establish the true multiplicity of \BAT, to characterise the orbit(s) of the system, to measure the physical properties of its individual components, and to determine the overall evolutionary status of the system.}
   {Using newly acquired high resolution spectra taken with the Ultra-violet and Visual Echelle Spectrograph mounted on the Very Large Telescope, we measured radial velocities via cross-correlation and line-profile fitting and performed a spectral analysis of the individual components using model atmosphere codes. 
   We estimated the age of the system and derived an evolutionary scenario for the 1.55-day system
   .}
   {\BAT\space comprises at least four components. The 1.55-day photometric signal originates in an eclipsing binary that consists of two O-type stars of spectral types O4\,V and O6.5\,V, which are both rapid rotators ($295$\,km\,s$^{-1}$ and $210$\,km\,s$^{-1}$, respectively). From the broad emission lines of the WR star, we derived a spectral type WN2.5-3. We further reject the previously reported 25.5-d period for the WR star and find that there is no detectable orbital motion within our uncertainties. The presence of additional narrow Si\,{\sc iii} and O\,{\sc ii} lines in the composite spectrum corresponds to a fourth component, a B1\,V star. There is clear evidence that the B-type star shows a radial velocity variation; however, the data do not allow for a determination of the orbital parameters. The configurations of the B-type star, the WR star, and possible additional undetected components remain unknown. We derived masses for the O-type components of $36\pm5\Modot$ and $15\pm2\Modot$, respectively, and estimated the age of the system to be 4.2\,Myr. We find evidence of previous or ongoing mass-transfer between the two O-type components and infer initial masses of $23\Modot$ for the O4\,V star and $29\Modot$ for the O6.5~V star. The O+O binary likely went through a phase of conservative mass transfer and is currently a near-contact system.}
   {We show that \BAT\space is a multiple -- quadruple or higher-order -- system with a total initial mass of at least 160~$M_\odot$. The 1.55-day O+O binary most likely will not evolve towards a BH+BH merger, but instead will merge before the collapse of components to BHs.}


   \maketitle
%

\section{Introduction}

Massive stars ($\Mini \gtrsim 8\Modot$) exhibit intense radiation fields and powerful stellar winds. Together with their final explosion as supernovae, they ionise the gas and influence the star formation rates in their surroundings, and they drive the chemical evolution of their host galaxies. Born as O- and early B-type stars, they end their life as compact objects -- neutron stars or black holes (BHs). Those that are massive enough are expected to first transition through a Wolf-Rayet (WR) phase, which is characterised by powerful stellar winds and broad emission line spectra \citep[e.g.][]{Crowther_2007}.\nextline
 Wolf-Rayet stars corresponding to the late stage of the evolution of massive stars (so-called classical WRs) are hydrogen-depleted and are categorised in three sequences: the nitrogen sequence (WN), the carbon sequence (WC), and the rare oxygen sequence (WO), which is based on whether their atmosphere is nitrogen-, carbon-, or oxygen-rich. It is thought that these three classes form an evolutionary sequence: WN $\rightarrow$ WC $\rightarrow$ WO \citep{Crowther_2007}. 


Broadly speaking, there are two WR-formation channels.
On the one hand, a single massive star can lose its hydrogen-rich envelope as a result of stellar winds or outbursts \citep{Conti1983}. On the other hand, the hot interior of a massive star can be revealed when the donor in a binary system is stripped from its outer hydrogen envelope by mass-transfer \citep[e.g.][]{Paczynski_1967}. In both scenarios, the stripped star then develops a strong radiation-driven wind, leading to the formation of a WR star. The relative importance of either channel remains however an ongoing debate \citep[e.g.][]{Vanbeveren1998, Neugent_2014, Shenar_2020}. In this context, establishing the evolutionary status of WRs in binary systems is of particular interest as it may offer ways to better constrain the impact of multiplicity on its evolutionary history and, from there, refine our physical understanding of binary evolution. 
Several studies \citep[e.g.][]{Sana_2014, Moe_2017} have indeed shown that over 90\% of massive stars are part of binaries or higher-order multiple systems and that most of them interact with a companion during their evolution \citep{Sana_2012, Sana_2013}. 

Furthermore, the study of WR multiple systems is also relevant in the context of the gravitational-wave events that result from the coalescence of stellar-mass BHs \citep[e.g.][]{Abbott_2016b, Abbott_2016a, Abbott_2017}. 
Indeed, one of the scenarios leading to these BH-merger events  involves the evolution of metal-poor massive binary stars with orbital periods below $3$\,days. In such systems, tidal synchronisation operates on a timescale that is short compared to the main-sequence lifetime \citep{Zahn_1975}, resulting in rapid rotation. As shown by \citet{de_Mink_2009}, efficient mixing induced by rotation can cause a chemically homogeneous evolution (CHE) of the primary star. This prevents the star from expanding, resulting in BH formation without previous binary interaction despite the short period. 
\nextline 
The further evolution then depends on the mass-ratio. In binary systems with initial mass ratios close to unity, both stars are expected to mix efficiently and form BHs, leading to a short period binary BH system that can merge within a Hubble time \citep{Mandel_2016, Marchant_2016}.
Alternatively, in systems with an initial mass ratio of $M_2/M_1<0.5$, only the more massive star will undergo CHE. The lower mass component is not massive enough to undergo CHE and it will expand as it evolves. The secular expansion of the lower mass component is then likely to lead to a phase of mass transfer between the companion star and the BH remnant of the initially more massive star. This phase is observed as an ultra-luminous X-ray source (ULX, $L_{\text{X}}>10^{39}$\,ergs; \citeauthor{Marchant_2017}, \citeyear{Marchant_2017}), which are sources commonly observed in star forming regions \citep[see][for a recent review on ULXs]{Kaaret_2017}. 
\nextline
BAT99\,126 is an eclipsing WR multiple system located in the Large Magellanic Cloud (LMC). It has been previously classified as a WN4+O8 binary with an orbital period of 25.5\,days \citep{Foelmi_2003b}. In addition, a previous light curve (LC) reveals an eclipsing system with a period of 1.55\,days \citep{Graczyk_2011}. The combination of the two periods suggests that BAT99\,126 might be a triple system \citep{Shenar_2019}. The configuration of the WR star and a close 1.55-day binary makes \BAT\space an excellent candidate to study close-binary interactions and hierarchical-system evolution.\nextline
Here, we perform a comprehensive spectroscopic and photometric analysis of the massive multiple system BAT99\,126 using new high-resolution spectroscopic data as well as archival photometric data, in order to constrain the configuration of this system and determine whether the 1.55-day binary could be the progenitor of a BH merger or ULX. We describe the observations and their reduction in \Secref{sec_data} and identify the components in \Secref{sec_components}. 
Section \ref{sec_orbital_analysis} presents the orbital analysis of the 1.55-day binary using the existing LC and the newly derived radial velocities. Section \ref{sec_spectral} discusses the classification of the individual components. Finally, in \Secref{sec_evolution}, 
we perform simulations on the evolution of the 1.55-day binary using the stellar evolution code MESA \citep[Modules for Experiments in Stellar Astrophysics;][]{Paxton_2011, Paxton_2013, Paxton_2015, Paxton_2018, Paxton_2019}, to investigate the previous and future evolution of the system. We end with a summary in \Secref{sec_summary}. 

\section{Observations and data reduction}\label{sec_data}

\begin{figure*}
    \centering
    \includegraphics[width = \textwidth]{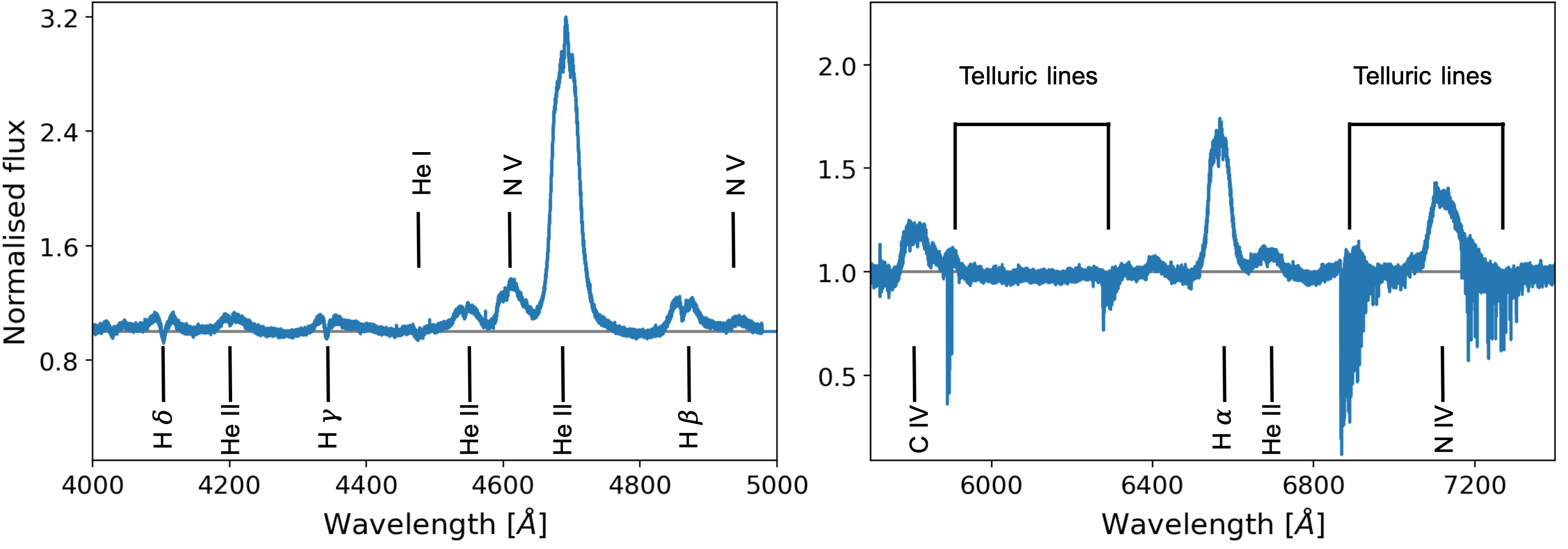}
    \caption{Concatenated normalised UVES spectrum taken at MJD = 58394.29 ($\phi_{\text{1.55d}}=0.32$). Left: UVES blue arm from 4000\,\AA\space to 5000\,\AA\space. Right: UVES red arm from 5700\,\AA\space to 7300\,\AA.}
    \label{figure_data_normalised_spectrum}
\end{figure*}

\subsection{Optical spectroscopy}
The spectroscopic data consists of ten spectra taken in 2018 over a 2-month period with UVES \citep[Ultra-violet and Visual Echelle Spectrograph;][]{Dekker_2000} mounted on the ESO Very Large Telescope\footnote{ID: 0102.D-0050(A), P.I.: T. Shenar}. Each spectrum is the result of three subsequent exposures obtained with the DIC2 437+760 setup, providing a wavelength coverage between 3755\,-\,5000\,\AA\space and 5700\,-\,9460\,\AA\space respectively. Each individual observation was taken with an exposure time of 1200 seconds and a slit width of 1.4" delivering a resolving power $R\approx32\,000$. The observing campaign was designed to cover the 1.55-day period, but given that it spans a few months, it also provides a fair coverage of the reported 25.5-day period. The journal of the observations and the corresponding phases of the known orbits
are given in \Tabref{table_data_UVES_observations}. \nextline
The data were reduced with the UVES pipeline, resulting in a 1D spectrum at each epoch. 
The consecutive observations are co-added by calculating the mean at every wavelength point, resulting in spectra with a signal-to-noise ratio (S/N) between 95 and 120 for the blue arm and between 45 and 65 for the red arm. 
We normalised the blue and red part of the co-added spectra separately. For this purpose, we used appropriate spectral energy distributions (SED) calculated with the Potsdam Wolf-Rayet (PoWR) code (\citeauthor{Hamann_2003} \citeyear{Hamann_2003}, see \Secref{sec_spectral}), scaled and reddened to normalise the spectra in a smooth, unbiased way. The reddening is accounted for as presented by \citet{Howarth_1983}. An example of a concatenated normalised spectrum is shown in \Figref{figure_data_normalised_spectrum}, clearly revealing  strong and broad emission line characteristics of WR stars. 

\begin{table}
\renewcommand{\arraystretch}{1.1}
\centering
\caption{Journal of the UVES observations of \BAT. The second and third columns give the phases corresponding to the 1.55-day and 25.5-day periods.}
\begin{tabu}{ crr}
     \hline
     \hline
     MJD$^{\text{a}}$ & \textcolor{white}{b}$\phi_{\text{1.55d}}$ $^{\text{b}}$ &\textcolor{white}{b}$\phi_{\text{25d}}$ $^{\text{c}}$\\ 
     \hline
     58394.29& 0.32 & 0.05\\
     58411.23 & 0.23 & 0.72\\
     58413.24 & 0.52 & 0.80\\
     58437.31 & 0.03 & 0.74\\
     58438.22 & 0.61 & 0.77\\
     58444.10 & 0.40 & 0.00\\
     58446.09 & 0.68 & 0.08\\
     58446.25 & 0.78 & 0.09\\
     58448.28 & 0.09 & 0.17\\
     58457.29 & 0.89 & 0.52\\\hline

\end{tabu}
    \begin{tablenotes}
      \small
      \item \textbf{Notes.} $^{\text{(a)}}$ Mid-exposure, in JD - 2\,400\,000.5; $^{\text{(b)}}$ MJD0 = 56\,454.16 (see \Secref{sec_orbital_Os}); $^{\text{(c)}}$ MJD0 = 51\,114 \citep{Foelmi_2003b}
    \end{tablenotes}
\label{table_data_UVES_observations}
\end{table}

\subsection{Photometric data}
The majority of the photometric data available for \BAT\space were obtained in the framework of the Optical Gravitational Lensing Experiment (OGLE; \citeauthor{Udalski_1992} \citeyear{Udalski_1992}; for the OGLE-IV catalogue see \citeauthor{Pawlak_2016} \citeyear{Pawlak_2016}; OGLE catalogue ID: OGLE-LMC-ECL-22000), providing over 800 I-band data points taken in a time span of about 13 years (October 2001 - March 2014). \nextline
The OGLE data yield a clear photometric period of 1.552945\,days \citep{Graczyk_2011}. Other photometric data include XMM-OT UV and optical photometry  \citep{Page_2012}, UBVRI photometry \citep{Bonanos_2009} as well as IR photometry from VISTA \citep{Cioni_2011}, 2MASS \citep{Zacharias_2013}, WISE \citep{Cutri_2012}, and Spitzer \citep{Meixner_2006}. These are summarised in Appendix \ref{appendix_photometric_data}.

\subsection{Imaging}\label{sec_imaging}
\begin{figure}
    \centering
    \includegraphics[width = 0.5\textwidth]{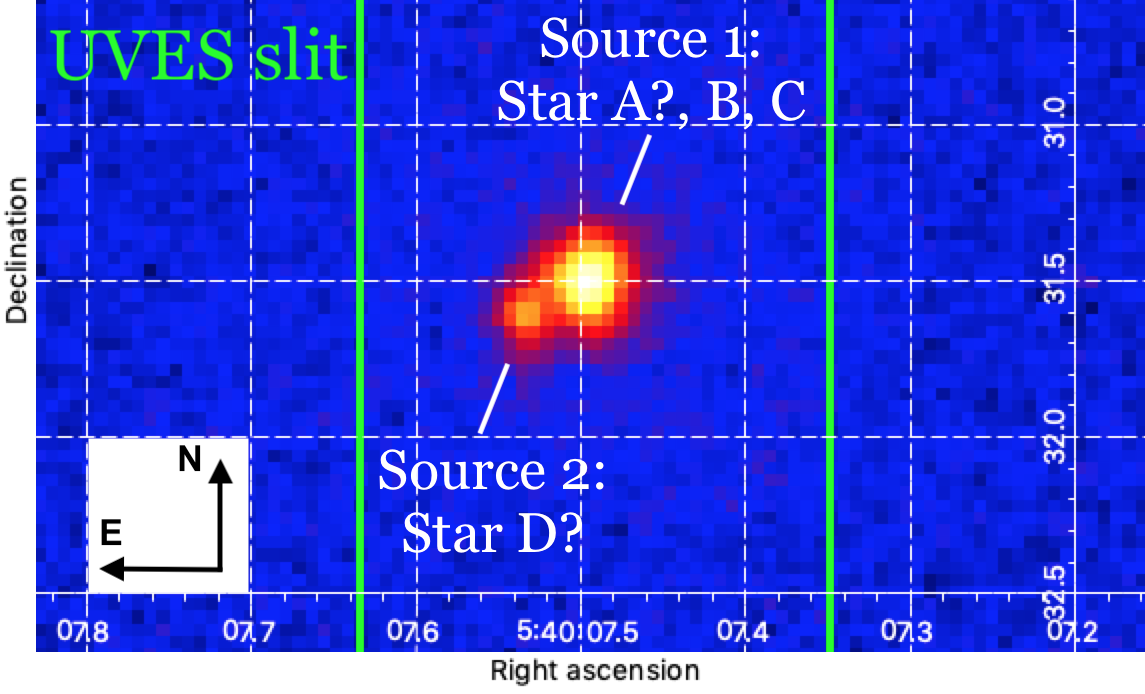}
    \caption{Hubble Space Telescope UV WFC3/UVIS image of \BAT, with an overlay of the UVES slit (width = 1.4"). Two well-separated sources are clearly visible, for which the likely corresponding components are indicated. North is up, east to the left.}
    \label{figure_HST}
\end{figure}

\BAT\space was also observed by the Hubble Space Telescope (HST), yielding a high-angular resolution image of the system. The image is a 2-second exposure taken with the WFC3/UVIS (Wide Field Camera 3/Ultraviolet-Visible) instrument and the filter F225W in the ultra-violet (UV) wavelength band\footnote{ObsID: hst\_12940\_10\_wfc3\_uvis\_f225w, PI: P. Massey}, covering a field-of-view of 2.7 $\times$ 2.7 arcmin. A zoom-in on \BAT\space is shown in \Figref{figure_HST} with the UVES slit indicated. The presence of two distinct stellar sources is readily seen, a brighter one in the north-west and a fainter one in the south-east. The separation between the two sources is $\sim\,$20\,mas corresponding to $\sim\,$10\,000\,AU at the LMC distance \citep[49.97\,kpc,][]{Pietrzynski_2013}. Both sources are located well within the UVES slit, meaning that if the sources are bright enough they are visible in the spectra. We estimated the flux of the sources by fitting their point-spread functions using a \textsc{python} code based on the \textsc{photutils}\footnote{\url{https://photutils.readthedocs.io}} package. We find that the flux ratio between the fainter and brighter light sources is about 1/10.


\section{The components of \BAT}\label{sec_components}

Because of the high-order multiplicity of \BAT\space and the resulting complexity of the spectrum, we first provide an overview of the spectral signature of each detected component. These are substantiated by the different panels in \Figref{fig_multiplicity}, where the bottom-right panel shows a schematic overview of the configuration of \BAT\space.

\subsection{Component A (WN2.5-3)}
\begin{figure*}[h!]
    \centering
    \begin{subfigure}{0.492\linewidth}
    \includegraphics[width =1\textwidth]{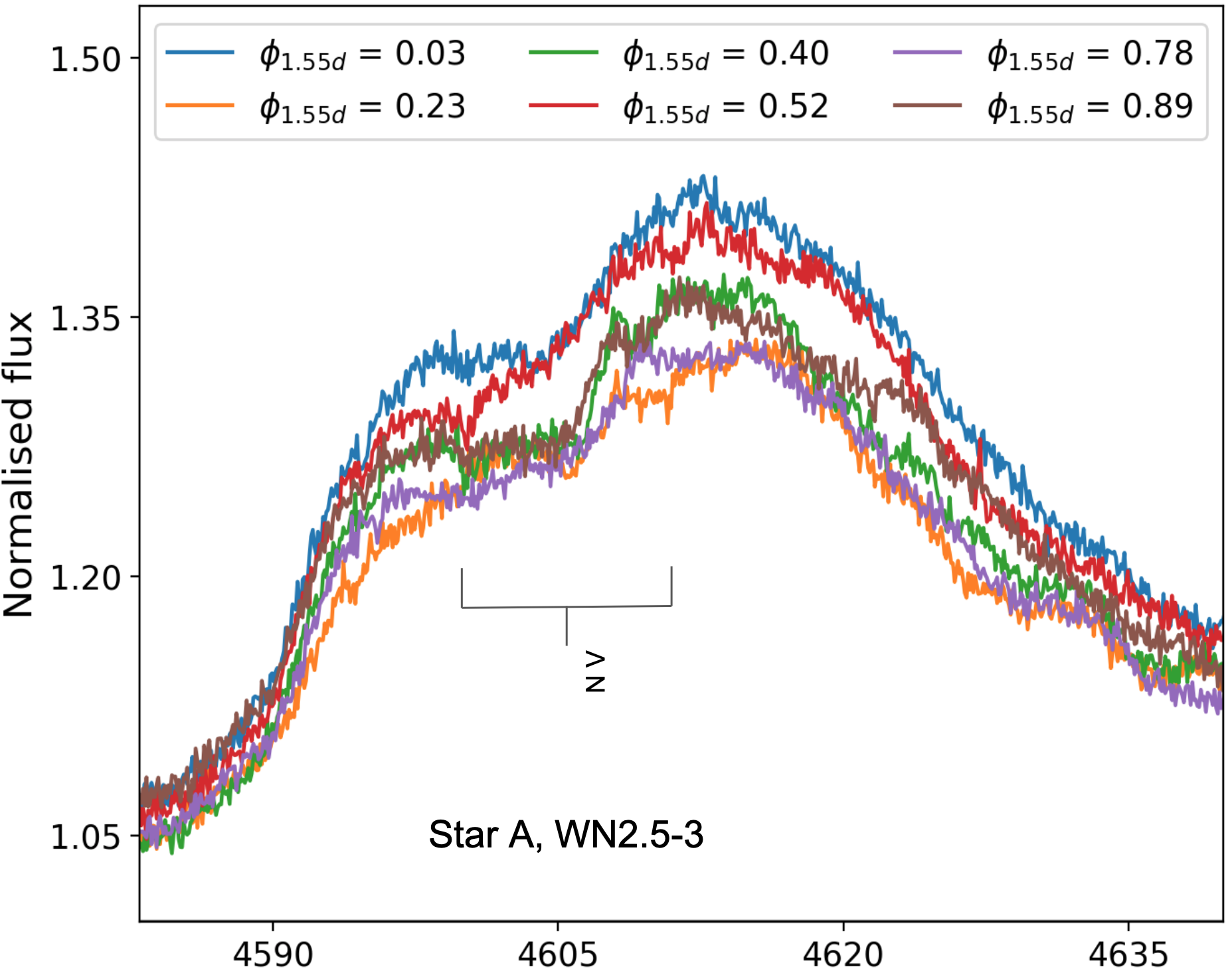}
    \end{subfigure}
    \hfill
    \begin{subfigure}{0.492\linewidth}
    \includegraphics[width =1\textwidth]{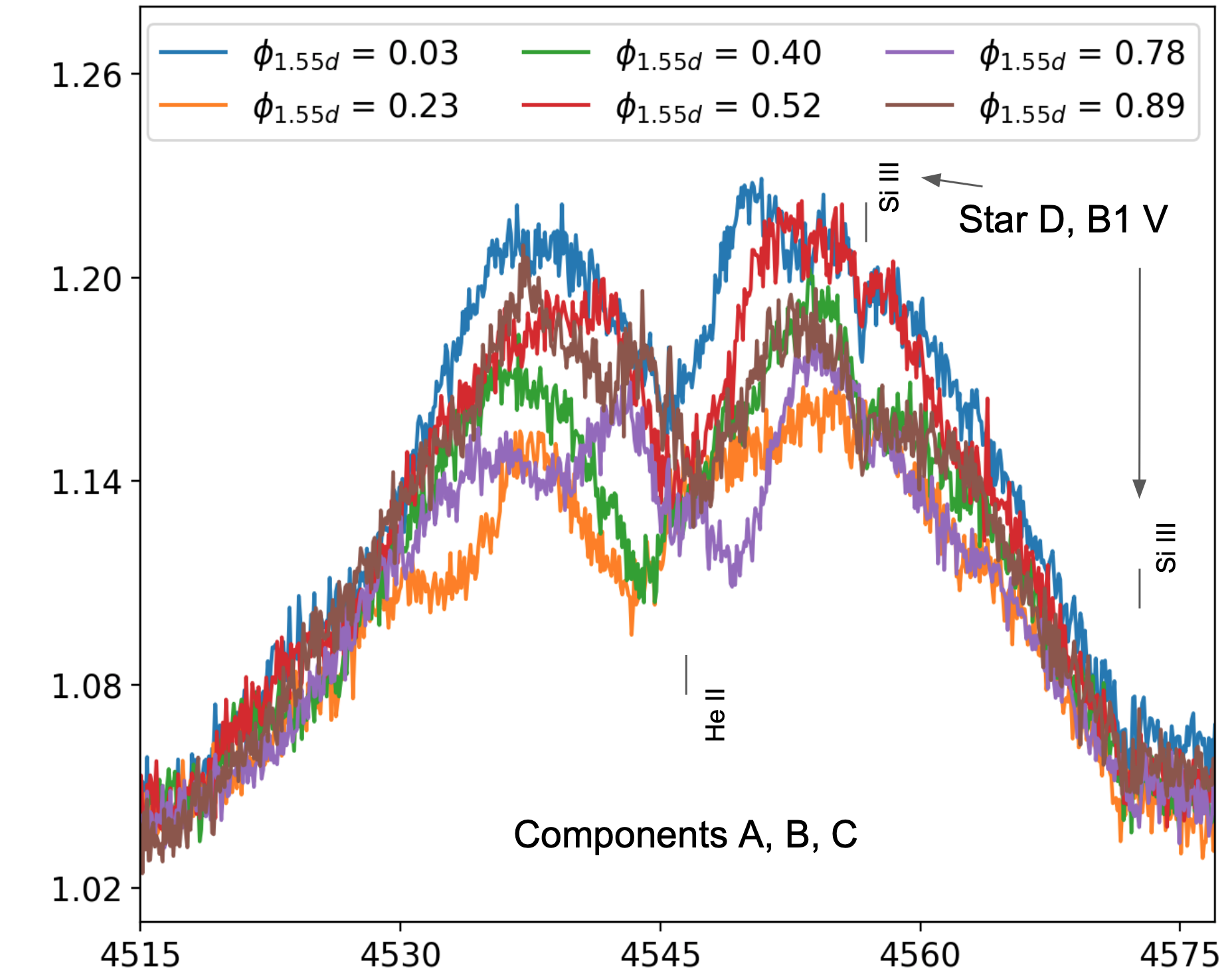}
     \end{subfigure}
   \vspace{-0.4cm}
    \begin{subfigure}{0.492\linewidth}
    \includegraphics[width =1\textwidth]{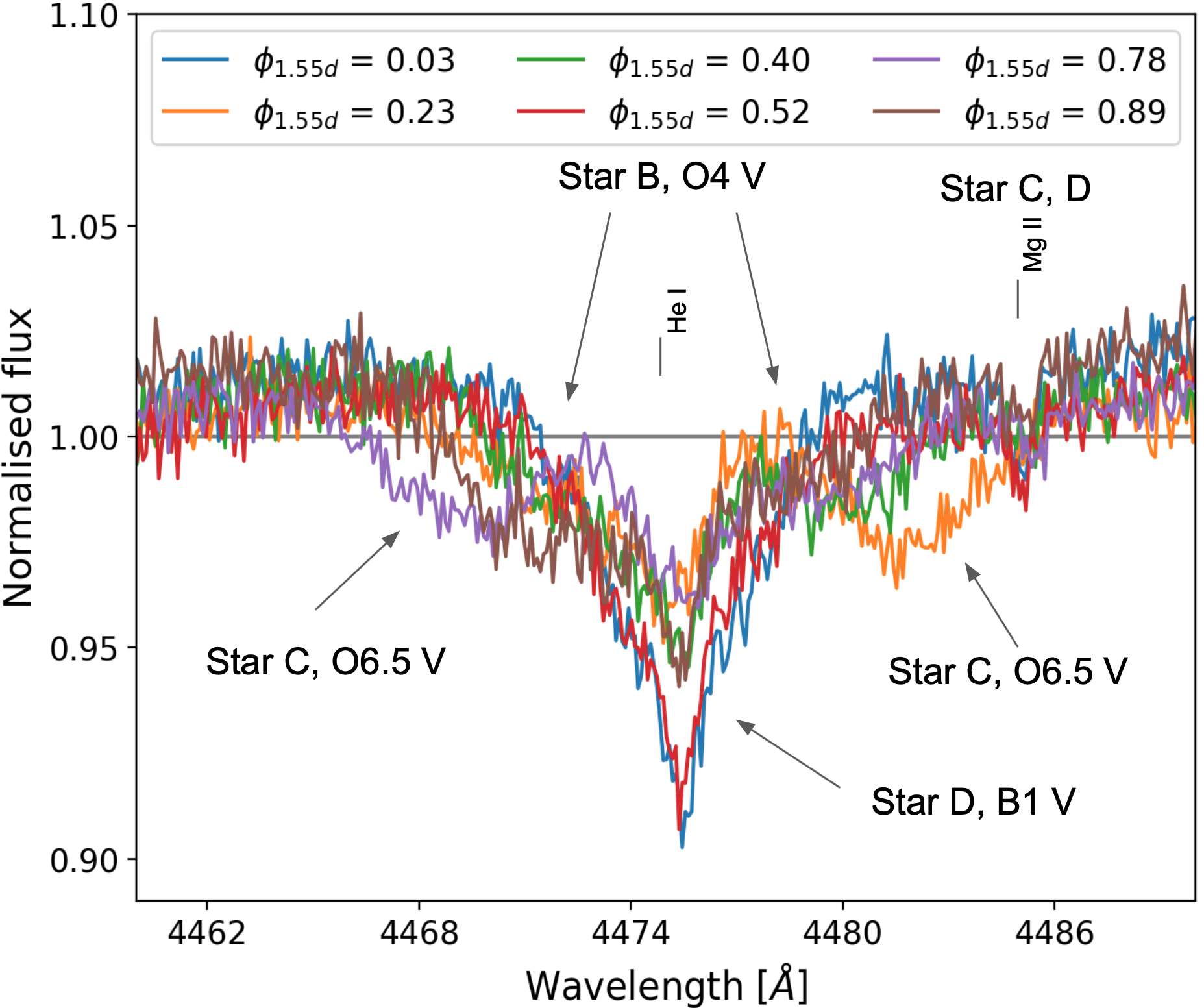}
    \end{subfigure}
    \hfill
    \begin{subfigure}{0.492\linewidth}
    \includegraphics[width =1\textwidth]{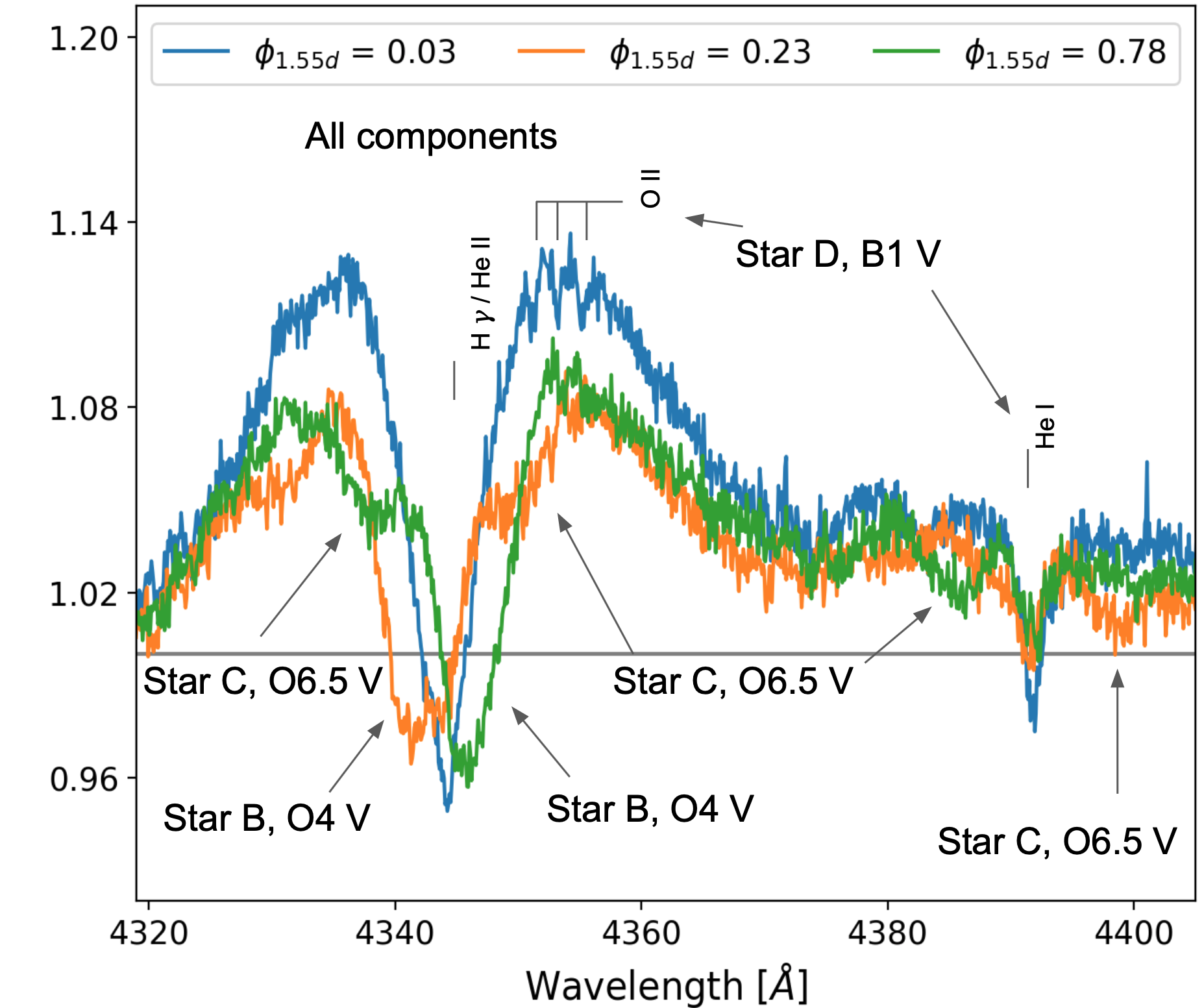}
    \end{subfigure}
    \begin{subfigure}{0.492\linewidth}
    \includegraphics[width =1\textwidth]{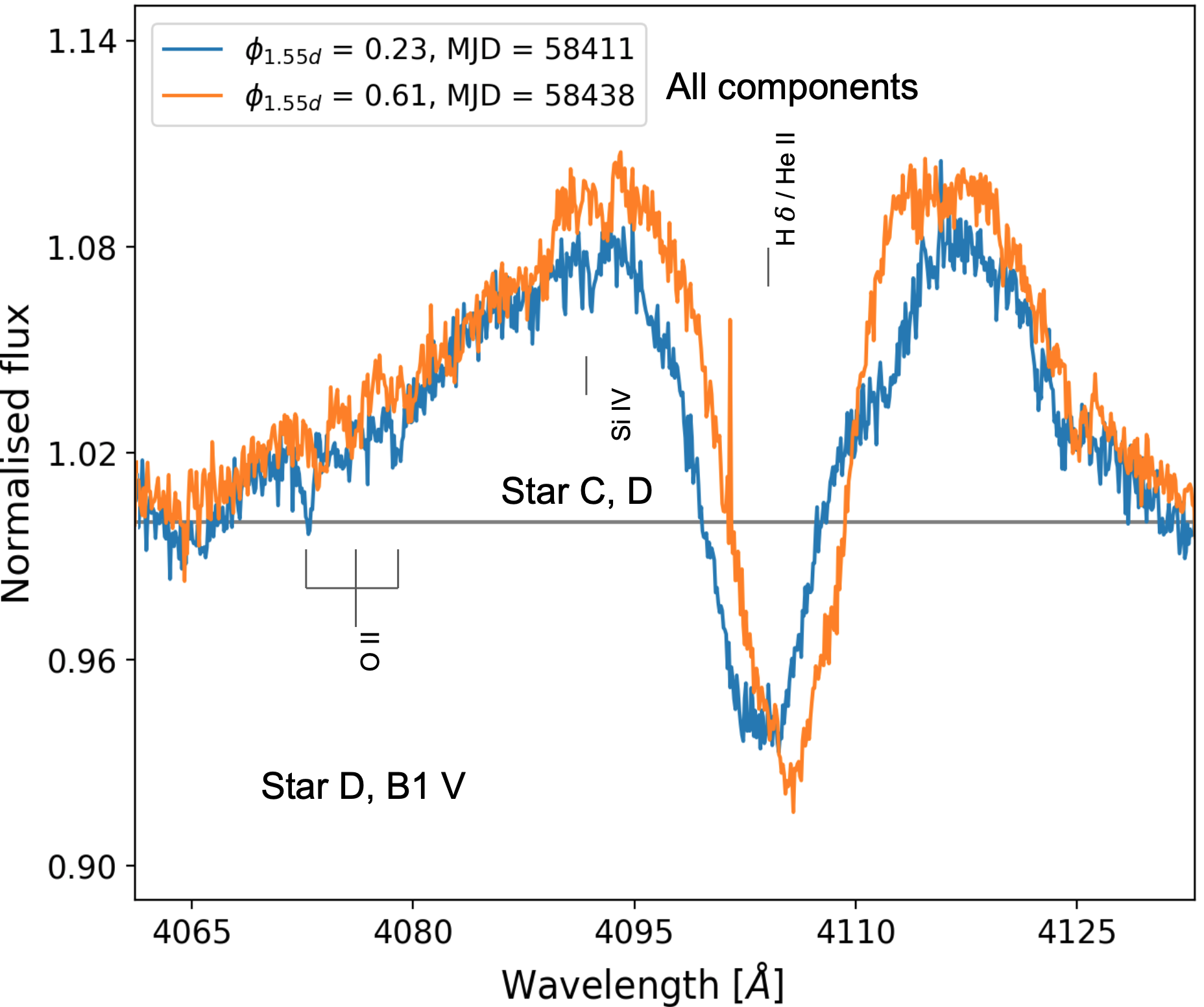}
    \end{subfigure}
    \begin{subfigure}{0.492\linewidth}
    \includegraphics[width =1\linewidth]{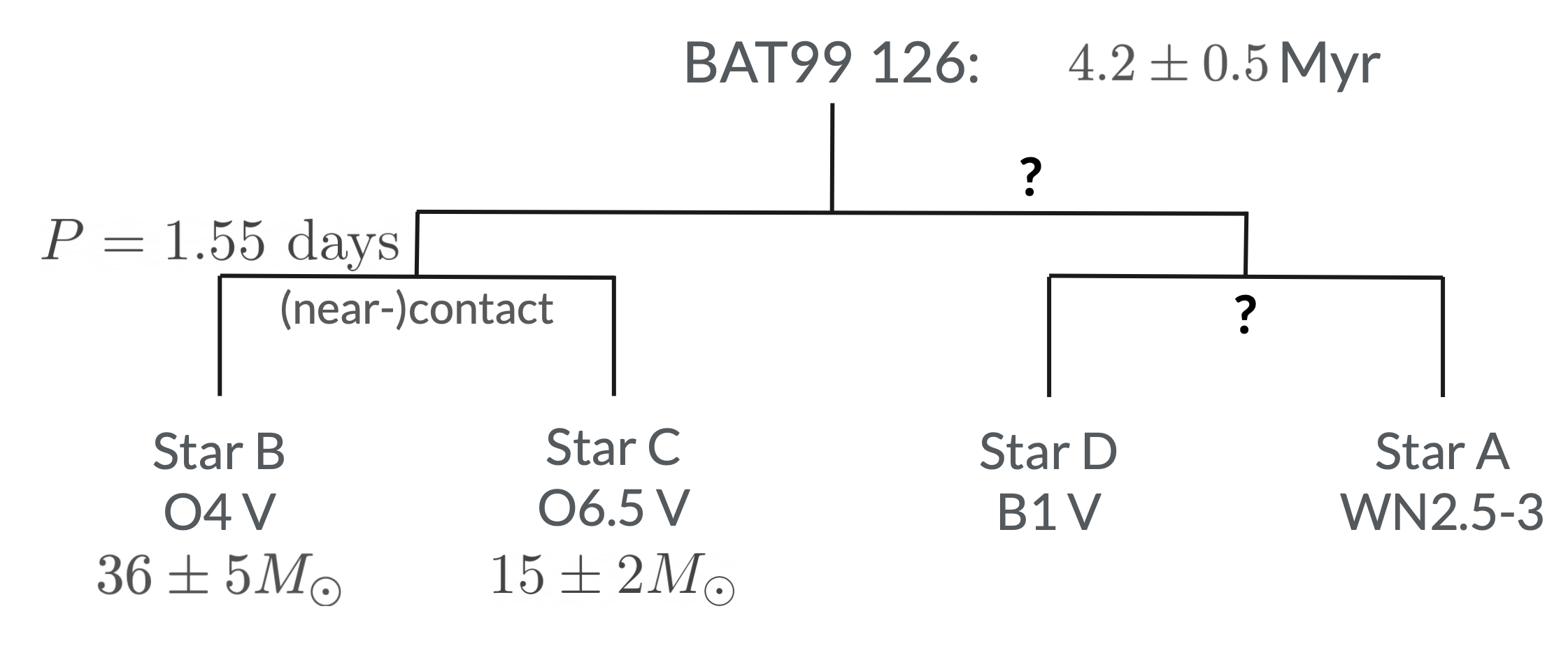}
    \end{subfigure}
    \caption{Featured spectral lines, illustrating the four components and their motion in the first five panels (see labels and legend). These UVES spectra are binned to 0.1\,\AA\space for clarity. The bottom-right panel shows a schematic overview of the configuration of \BAT.}
    \label{fig_multiplicity}
    \label{fig_scheme}
\end{figure*}

The WR star (WN2.5-3, see \Secref{sec_spectral_types_teff}) is easily distinguishable because of its broad emission lines (e.g. N\,{\sc v}\,$\lambda\lambda$4604, 4620 and He\,{\sc ii}\,$\lambda$4542, top-left and -right panels of \Figref{fig_multiplicity}). While \citet{Foelmi_2003b} reported that the WR star participates in a 25.5-day orbit, no clear orbital motion is seen in the UVES data (see \Secref{sec_orbital_ana_WR+B}).

\subsection{Components B (O4\,V) and C (O6.5\,V)}
The spectra of \BAT\space show two broad absorption features in the He\,{\sc ii}, He\,{\sc i}, and the Balmer lines that are moving in anti-phase. These features belong to two additional components, B and C (O4\,V and O6.5\,V respectively, \Secref{sec_spectral_types_teff}). 
Component B is most prominent in He\,{\sc ii} and hydrogen lines (e.g. He\,{\sc ii}\,$\lambda$4542, top-right panel of \Figref{fig_multiplicity}), but is also weakly present in He\,{\sc i} lines. Component C can be seen in the He\,{\sc ii} lines, but it can be more clearly distinguished in the He\,{\sc i} lines (e.g. He\,{\sc i}\,$\lambda$4471, middle-left panel of \Figref{fig_multiplicity}). The anti-phase motion of these two components is highlighted in the middle-right panel of \Figref{fig_multiplicity}, which shows the H$\gamma$ line in one eclipsing spectrum and both the quadrature spectra. It can also be seen here that the hydrogen lines of component B are stronger than those of component C. The broadness of the spectral features of both components indicate they are fast rotators.\nextline
The semi-amplitudes of the radial velocity (RV) variations of components B and C are relatively high ($\approx 150$ and $\approx 400$\,km\,${\rm s}^{-1}$ respectively, see \Secref{sec_orbital_Os}) and suggest a mass ratio close to $q=M_{\text{C}}/M_{\text{B}}\sim0.4$. The two components are associated with the 1.55-day period seen in the LC (\Secref{sec_orbital_Os}).

\subsection{Component D (B1\,V)}
In addition to the spectral lines of the other three components, the spectra of \BAT\space show narrow absorption features. Clear examples are shown in \Figref{fig_multiplicity}, in the top-right panel for Si\,{\sc iii}\,$\lambda\lambda$4553, 4568 and the bottom-left panel for the O\,{\sc ii}\,$\lambda\lambda$4069, 4072, 4076 line. We also detect these narrow features in the He\,{\sc i} lines, but not in the He\,{\sc ii} lines (e.g. He\,{\sc ii}\,$\lambda$4542, top-right panel of \Figref{fig_multiplicity}). These spectral features are characteristics of a cooler ($\Teff < 32$\,kK) star and hence do not belong to the WR star or components B or C. Moreover, the absorption lines are too narrow to be coming from components B and C and they do not follow their Doppler-shifted motion. This fourth component, component D (B1\,V, see \Secref{sec_spectral_types_teff}), shows RV variability during the two months in which the spectra were taken, as can be readily seen in the bottom left panel of \Figref{fig_multiplicity} (see also \Figref{figure_motion_B}). However, no clear period is found (\Secref{sec_orbital_ana_WR+B}).

\section{Orbital analysis}\label{sec_orbital_analysis}

\subsection{Radial velocities}
In order to obtain the RVs of the individual components in the ten available UVES spectra, two techniques were used: cross-correlation and line-profile fitting.
The obtained RVs of the components are listed in \Tabref{table_RVs}.
\subsubsection{Cross-correlation (stars A and D)}
For components A and D (WN2.5-3 and B1\,V respectively, \Secref{sec_spectral_types_teff}), we adopted a technique similar to the one utilised in \citet{Shenar_2017, Shenar_2018}. The method cross-correlates a template (e.g. a model spectrum) with each individual spectrum \citep{Zucker_1994}. For spectral lines which are dominated by one component, the individual spectra can be co-added in the frame of reference of that component to create a master template with a higher S/N than a single spectrum. In order to do so, a first estimate of the RVs is necessary.\nextline
For the WR star, the first estimate of the RVs was obtained by using one of the ten available UVES spectra as the template. In particular, the RVs were first determined for the N\,{\sc v}\,$\lambda\lambda$4604, 4620 lines. Using the derived RVs, all ten spectra were co-added and we repeated the process with the co-added spectrum used as the template.
We determined RVs for the He\,{\sc ii}\,$\lambda$4542 and He\,{\sc ii}\,$\lambda$4686 lines and also for the N\,{\sc v}\,$\lambda\lambda$4604, 4620, C\,{\sc iv}\,$\lambda\lambda$5801, 5812 and N\,{\sc iv}\,$\lambda\lambda$7103, 7109, 7111, 7123, 7127, 7129 lines. The N\,{\sc v}, C\,{\sc iv} and N\,{\sc iv} lines result in more reliable RVs for two reasons. Firstly, the spectral lines coming from elemental transitions of higher ionisation stages are formed close to the stellar surface (e.g. N\,{\sc v} forms at $r\approx1.5R_{*}$). Therefore, the motion in these lines should represent the true motion of the WR star more accurately than motion derived from, for example, the He\,{\sc ii} lines. Secondly, the N\,{\sc v}, N\,{\sc iv} and C\,{\sc iv} lines only originate from the WR, hence they are not contaminated by the spectral signature of other components in this complex system. \nextline
The three different measurements for the RVs are not fully in agreement within errors, potentially due to the different formation regions of the lines. Hence, to obtain the RVs of the WR star, we calculated the average between the three data points (the aforementioned N\,{\sc v}, N\,{\sc iv} and C\,{\sc iv} lines) while the uncertainty on the average is taken as the standard deviation between the individual measurements.\nextline
The template used for component D (B1\,V, \Secref{sec_spectral_types_teff}) is a model spectrum retrieved from the B-star grid \citep{Hubeny1995, Lanz2007} calculated with TLUSTY, a plane-parallel model atmosphere code that relaxes the assumption of local thermodynamic equillibrium (non-LTE). We chose an appropriate model for a B1~V star with $\Teff = 25$\,kK and $\log g = 4.0$\,[cm s$^{-2}$] (\Secref{sec_spectral_types_teff}). The model was scaled to match a contribution of 10\% (\Secref{sec_spectral_contributions}) and a rotational broadening of 40\,km\,s$^{-1}$ (\Secref{sec_spectral_rotv}) was applied. The spectral lines used to determine the RVs of component D are Si\,{\sc iii}\,$\lambda$3807, Si\,{\sc iii}\,$\lambda$4553, Si\,{\sc iii}\,$\lambda$4568, O\,{\sc ii}\,$\lambda$3945, O\,{\sc ii}\,$\lambda\lambda$4069, 4072, 4076 and O\,{\sc ii}\,$\lambda\lambda$4415, 4417. Most of these lines are located in the wings of the broad WR-emission lines (e.g. the Si\,{\sc iii}\,$\lambda$4568 line, top-right panel \Figref{fig_multiplicity}). To remove possible motion and variability due to the WR star, we renormalised a small region around the absorption lines, then we followed the same RV measurement procedure as described above for the WR star.\nextline
To conclude, we note that the RVs of the WR star are relative to the co-added template, whereas the RVs of component D are measured against the rest wavelength of the atomic transitions. This means that we do not have a direct measurement of the systemic velocity of the WR star.

\begin{figure}
    \centering
    \includegraphics[width = 0.5\textwidth]{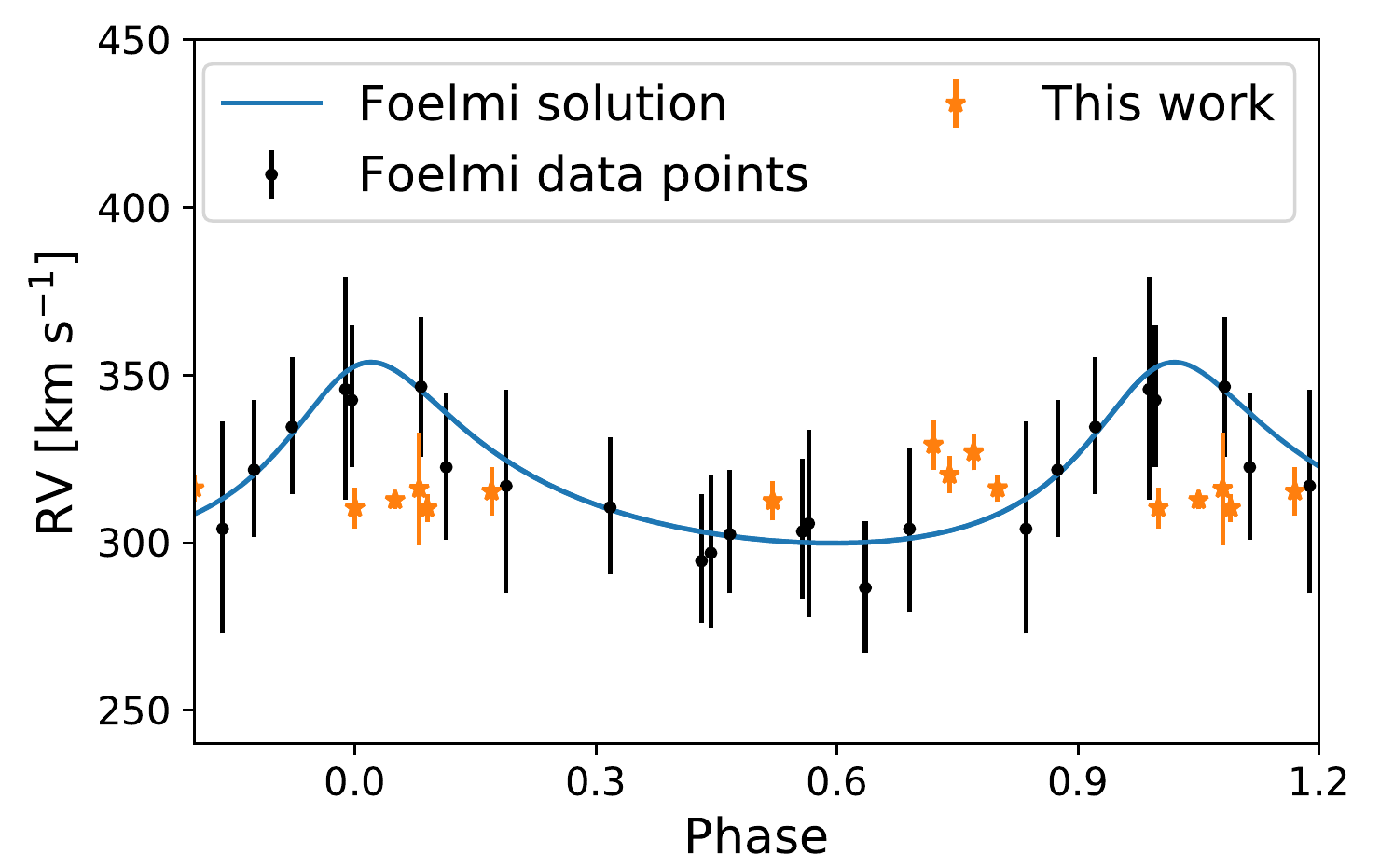}
    \caption{Radial velocities obtained for the WR star, marked with orange stars. The points are phased to the 25.5-day orbit derived by \citet{Foelmi_2003b}, which is shown by the blue curve. Their derived data points are also marked with black dots.}
    \label{figure_RVs_WR_Foelmi}
\end{figure}

\subsubsection{Line-profile fitting (stars B and C)}
The line-profile fitting approach that we adopted is similar to the one presented in \citet{Mahy_2010, Mahy_2011, Mahy_2017}. This technique uses Gaussian profiles to fit to the spectral lines, allowing for multiple components at once. Here, we used this method to derive the RVs of components B and C (O4\,V and O6.5\,V respectively, \Secref{sec_spectral_types_teff}), as all available spectral lines for these two stars are entangled with each other or those of the other two components.\nextline
For the RV determination of components B and C, we used He\,{\sc i}\,$\lambda$3820, He\,{\sc i}\,$\lambda$4471, He\,{\sc ii}\,$\lambda$4200 and He\,{\sc ii}\,$\lambda$4542. Both He\,{\sc i} lines are contaminated by star D and the He\,{\sc ii} lines are blended with the broad emission lines of the WR star. Components B, C and D were fitted simultaneously with Gaussian profiles, whereas the emission lines of the WR star were fitted with a Voigt profile. The full-width half-maximum (FWHM) and amplitude of all profiles were allowed to vary during the fitting as eclipses in the spectrum cause varying line strengths. Also the proximity of components B and C might cause the line shapes to vary at different phases of the orbit due to rotational distortions \citep{Micheal_2020}. The fitting was performed with the \textsc{python} \textsc{lmfit} package\footnote{\url{https://lmfit.github.io/lmfit-py/}}. \nextline
For the determination of the average RVs, we used the same method  as above for the WR star and component D. The RVs that are determined for the components B and C are also measured against the rest wavelength of the atomic transitions.

\subsection{On the orbital motion of the WR and the B-type star (stars A and D)} \label{sec_orbital_ana_WR+B}
We used the WR orbital period of 25.5\,days proposed by \citet{Foelmi_2003b} to phase-fold our own RV measurements and compared with earlier RV values\footnote{Since \citet{Foelmi_2003b} did not tabulate their RVs, we extracted them from their figure 3 with the online tool Graphreader: \url{http://www.graphreader.com}}. As previously mentioned, our measured RVs of the WR star are relative to the co-added mask. In order to provide a visual comparison of our data with the historical data of \citet{Foelmi_2003b}, we applied a systematic shift so that the average of our measurements corresponds to the systemic velocity derived by \citet{Foelmi_2003b}. Figure \ref{figure_RVs_WR_Foelmi} shows that our derived RVs do not agree with the 25.5-day orbit (no matter what systematic velocity shift we applied to the RVs). \nextline
As noted by \citet{Foelmi_2003b}, since their derived RV amplitude of $K=27\,$\kms\space is below their detection threshold, it is likely that their reported period is a spurious result. Their RVs were obtained by fitting Gaussians to the He\,{\sc ii}\,$\lambda$4686. However, this strong emission line is not well represented by a Gaussian. Moreover, the He\,{\sc ii}\,$\lambda$4686 line is more susceptible to variability than the N\,{\sc v}, N\,{\sc iv} and C\,{\sc iv} lines \citep[e.g.][]{Dsilva_2020}, which may explain the observed discrepancies with our results.\nextline

\begin{figure}
    \centering
    \includegraphics[width = 0.5\textwidth]{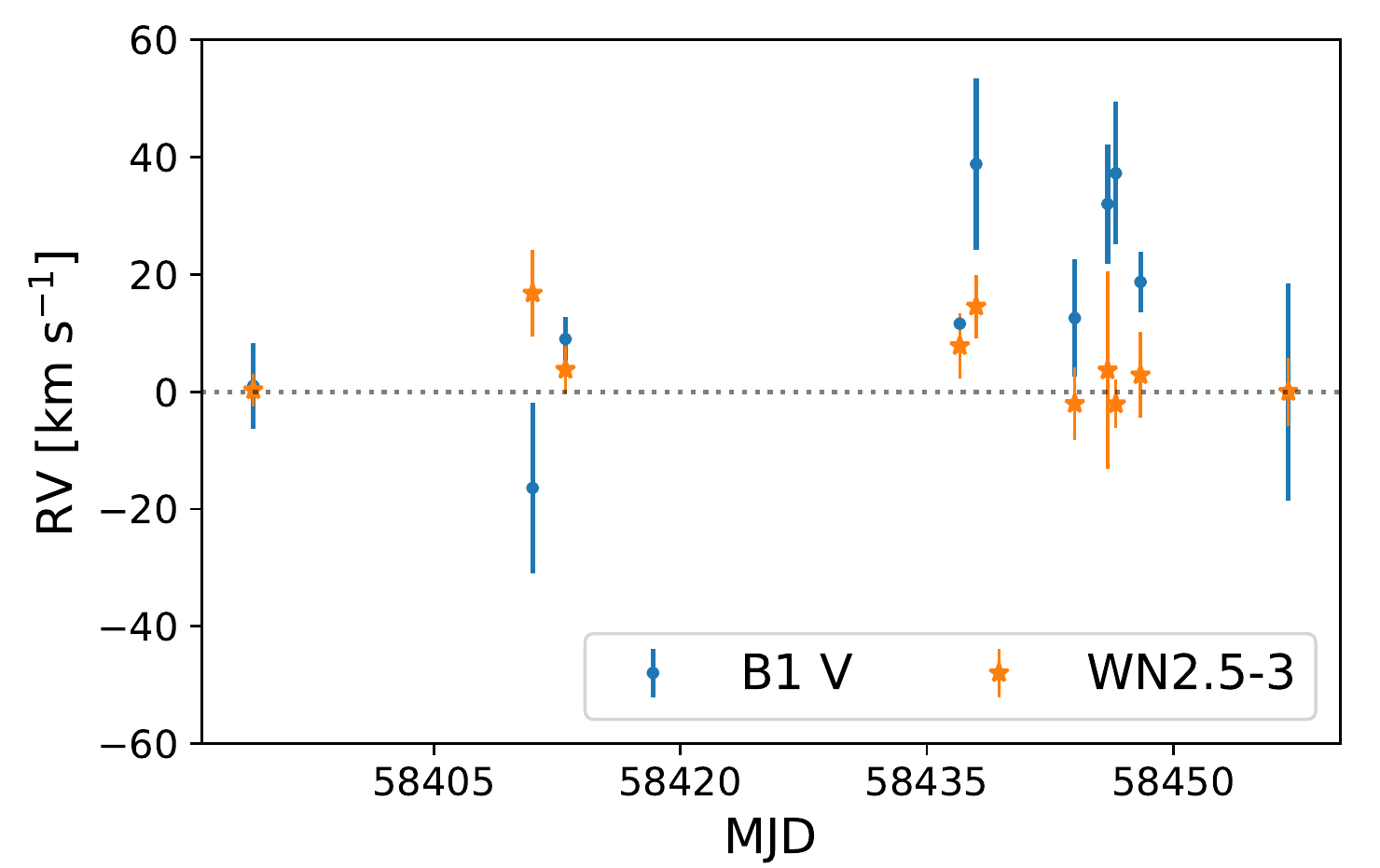}
    \caption{Radial velocities obtained for the WR star (orange stars) and component D (B1\,V, \Secref{sec_spectral_types_teff}) (blue dots) as a function of time when the spectrum was taken. Points are calibrated such that both stars have RV=0\,km\,s$^{-1}$ at the final data point.}
    \label{figure_RVs_WR+B}
\end{figure}

The relative RVs of the WR and the B stars are also displayed against the Julian date in \Figref{figure_RVs_WR+B}, adopting RV=0\,km\,s$^{-1}$ for both stars at the final epoch. Both the spectra and the obtained RVs show evidence for motion of component D (see also \Figref{figure_motion_B}). However, the data are too sparse to obtain an orbital period. It also remains unclear whether the WR star and component D are gravitationally bound to one another or whether component D is orbiting around an undetected companion, which would imply that \object{BAT99\,126} is a quintuplet. Based on our estimated flux contributions in the visual (see \Secref{sec_imaging}), we obtain an observed $V_{\text{faint}}$\,=\,15.8\,mag for the fainter source. This value is in good agreement with the average apparent $V$ magnitude of B1~V stars in the LMC of 15.93 mag obtained from the compilation by \cite{Bonanos_2009}, and is hence consistent with our derived spectral type. Therefore, we propose that
components A, B and C belong to the brighter source, while component D belongs to the fainter source. However, more data are required to confirm this (e.g. HST or MUSE (Multi Unit Spectroscopic Explorer at the VLT) - narrow-field mode). \nextline
If our proposed configuration is correct and if the RV variations seen in Fig.~\ref{figure_RVs_WR+B} are real, component D would then to be orbiting an undetected companion. A back-of-the-envelope calculation adopting a circular orbit, a period of 2\,months, and an orbital inclination $i=60$\degree, reveals that a $2.5\Modot$-object would be sufficient to produce the observed $\sim50$\,km\,s$^{-1}$\ peak-to-peak RV variations for component D if the latter has a mass $M_{\text{D}} = 10\Modot$. Such a low mass companion would indeed not be detectable in our UVES dataset. 

As no clear orbital motion is detected, it can also not be excluded that the WR star and B-type star are not bound to the 1.55-day system. However, the possibility that they are stars in the line-of-sight of the close binary is extremely unlikely, and it would be even more unlikely that this would the case for both stars.

\subsection{Orbital solution of the O+O binary (stars B and C)} \label{sec_orbital_Os}
\begin{figure}
    \includegraphics[width = 0.5\textwidth]{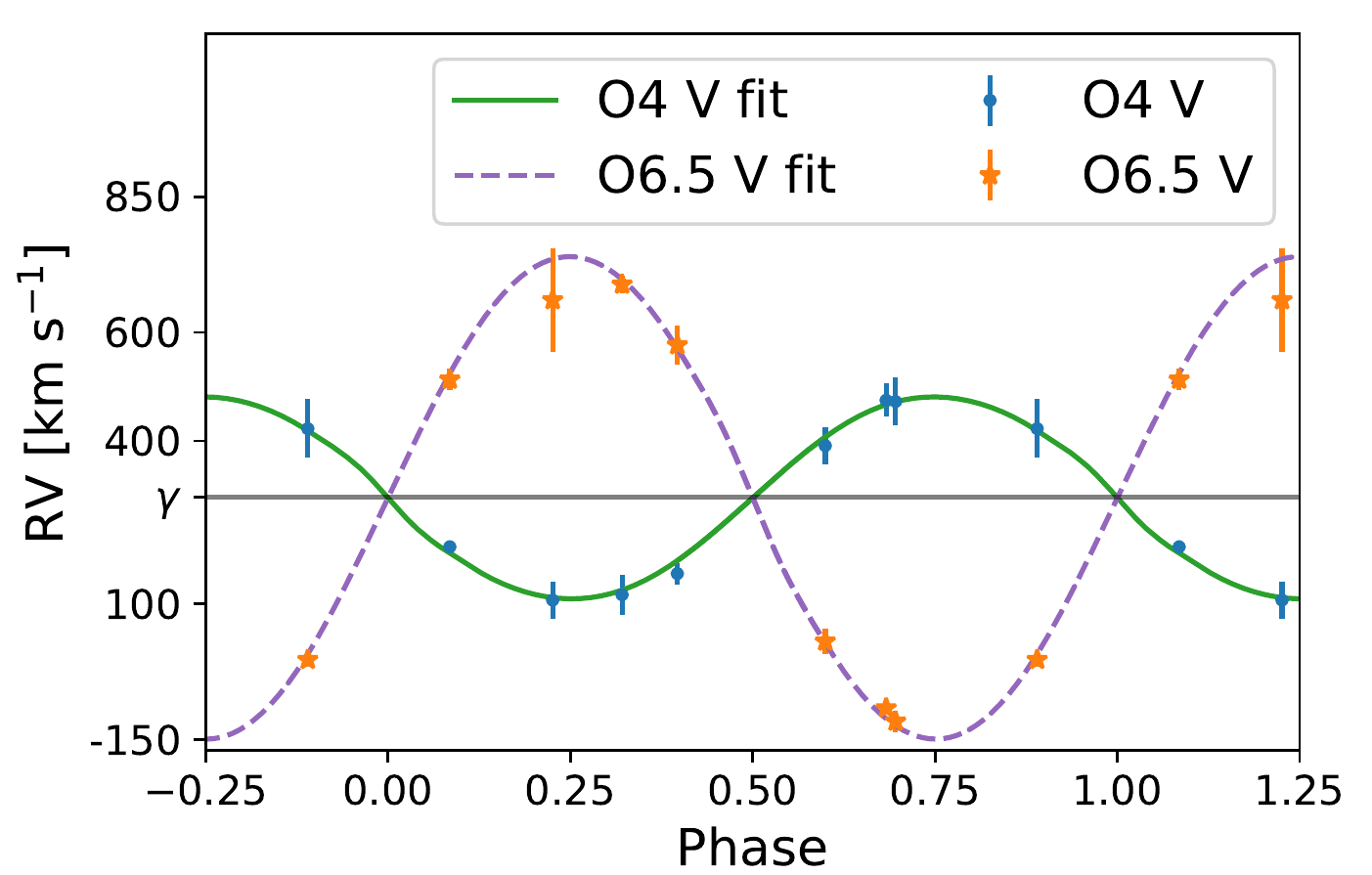}
    \caption{Radial velocity measurements and best-fit RV curves of the O+O binary according to the solution listed in \Tabref{table_orbital_param_Os}.
    The RVs at $\phi_{1.55\text{d}} = 0.03$ and $\phi_{1.55\text{d}} = 0.52$ are not included since the eclipses caused problems for the RV measurements.
    }
    \label{figure_RVs_Os}
\end{figure}

\begin{figure}
    \centering
    \includegraphics[width = 0.5\textwidth]{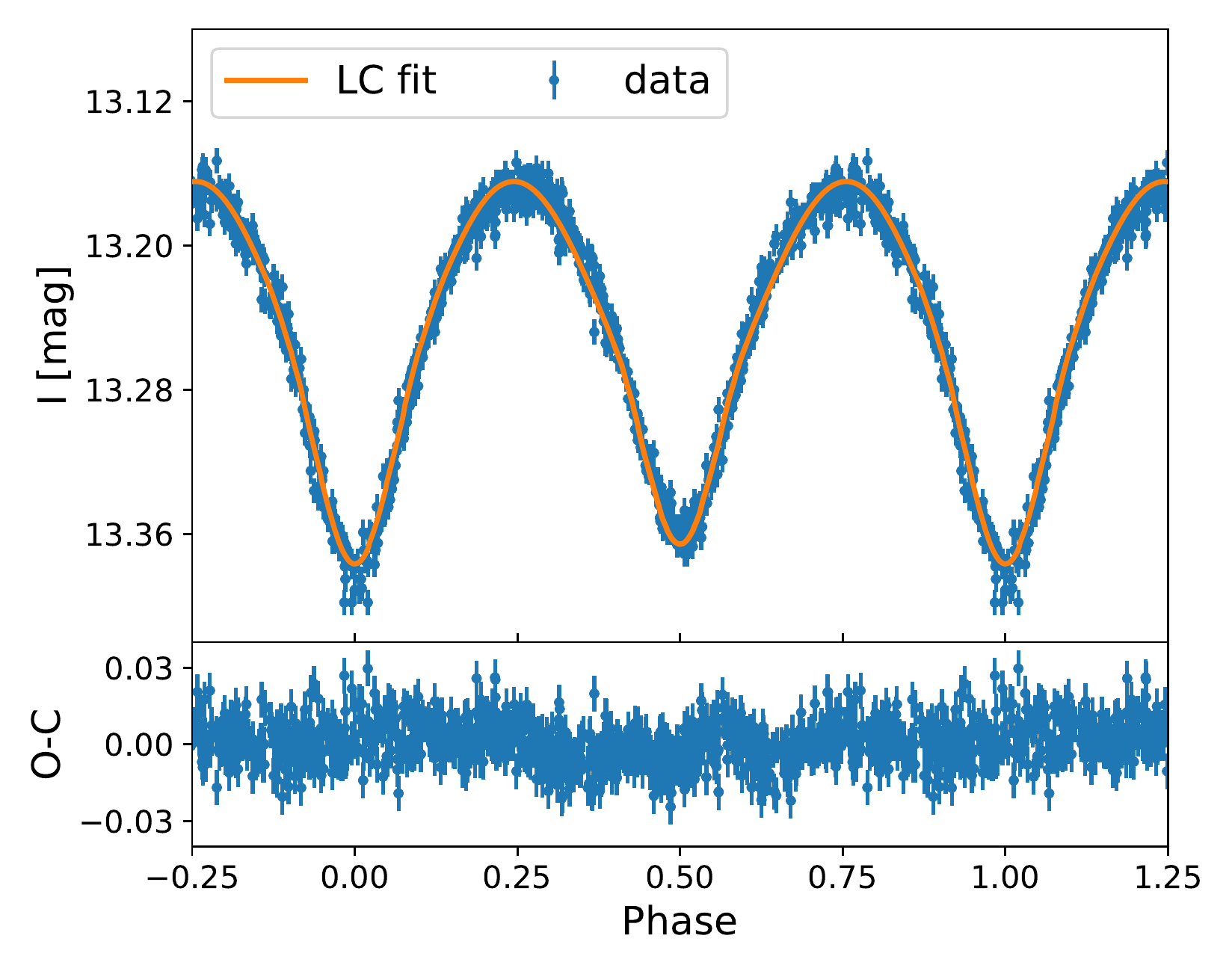}
    \caption{OGLE LC of \BAT\space phase-folded onto the 1.55-day orbit derived by \citet{Graczyk_2011} and the fit obtained from the orbital fitting. The bottom panel shows the residuals of the observed data minus the fit.}
    \label{figure_LC_Os}
\end{figure}

We combined our measured RVs with the OGLE light curve to perform an orbital analysis on the close O+O binary. For this, we used the Physics of Eclipsing Binaries (PHOEBE) LC modelling software \citep[][version 2.1]{Prsa_2005, PHOEBE_20, PHOEBE_21}, which fits synthetic light and RV curves to data, taking into account different orbital and stellar parameters. Uncertainties were derived according to the description in the PHOEBE manual\footnote{\url{https://usermanual.wiki/Document/phoebemanual.570010668/html}}. We fixed the period to the previously published period of 1.552945\,days \citep{Graczyk_2011}. Adopting the primary eclipse as phase $\phi = 0$, we constrained MJD0 = 56\,454.16. We also fixed the effective temperature of component B, hereafter called the primary, at 42\,500\,K (\Secref{sec_spectral_types_teff}), but left the effective temperature of the second O star (component C) as a free parameter. Component C  is hereafter referred to as the secondary in this O+O system.\nextline
The gravity darkening and surface brightness of both components are fixed to unity and we used a reflection effect with two reflections as we are dealing with a close binary with hot and massive components. We accounted for limb darkening using the square-root law, which is best used for components with $\Teff > 8\,500$\,K \citep{Diaz-Cordoves_1992}. \nextline
Given that the eclipses of the 1.55-day system are located at $\phi = 0.0$ and $\phi = 0.5$, we fixed the eccentricity to $e=0$. In order to take into account the contribution of the WR star and component D in the light curve, we finally adopted a third light contribution of 35\%. The fitting was repeated with values of 30\% and 40\%, without significant changes in the results.\nextline
\Tabref{table_orbital_param_Os} shows the obtained orbital and stellar parameters for the individual components of the short O+O binary. The corresponding RV curves are shown in \Figref{figure_RVs_Os} and the fitted light curve is shown in \Figref{figure_LC_Os}. It can bee seen from both figures that the obtained solution fits the data well. \nextline
We obtain an effective temperature of 38.0\,$\pm$\,1.9\,kK for component C, in excellent agreement with the value obtained from the spectroscopic analysis ($T_{\text{eff,\text{C},sa}}=37\,500\pm1\,250$\,K, \Secref{sec_spectral_types_teff}). Both $\log g$ values further agree with the $\log g$ value of a main-sequence star and with a value of $\sim4.0$ derived later by our spectral analysis. The critical value of the surface potential for both components is $\Omega(L_1) = 2.70$. Within the uncertainty on the surface potentials, it cannot be excluded that one or both stars are filling their Roche lobe.
Finally, using Stefan-Boltzmann's law $L = 4\pi R^2 \sigma \Teff^4$ we obtain for the primary $\log(L_\text{B}/L_{\odot}) = 5.41\pm0.22$ and for the secondary $\log(L_\text{C}/L_{\odot}) = 4.90\pm0.33$. 


\begin{table}
\centering
\caption{Orbital and stellar parameters for the 1.55-day binary, obtained from orbital analysis with PHOEBE. }
\begin{tabu}{ lr@{\hskip 0.03in}lr@{\hskip 0.03in}l}
     \hline
     \hline
     Parameter & \multicolumn{4}{c}{Value}\\ 
     \hline
     $P$ [d] & &\multicolumn{2}{c}{1.552945 (fixed)}&\\
     MJD0    & &\multicolumn{2}{c}{56\,454.16 (fixed)} &\\
     $e$ & &\multicolumn{2}{c}{\textcolor{white}{1.55294}0 (fixed)} &\\
     $q$ & &\multicolumn{2}{c}{\textcolor{white}{294}0.41 $\pm$ 0.03}  &\\
     $K_1$ [km s$^{-1}$] & &\multicolumn{2}{c}{\textcolor{white}{1.55}181 $\pm$ 10\textcolor{white}{0.}}& \\
     $K_2$ [km s$^{-1}$] & &\multicolumn{2}{c}{\textcolor{white}{1.55}440 $\pm$ 26\textcolor{white}{0.}}&\\
     $a \sin i$ [$R_\odot$] &  & \multicolumn{2}{c}{\textcolor{white}{94}19.06 $\pm$ 0.56}& \\
     $a$ [$R_\odot$] && \multicolumn{2}{c}{\textcolor{white}{94}20.95 $\pm$ 1.07} &\\
     $i$ [\degree] && \multicolumn{2}{c}{\textcolor{white}{94}65.49 $\pm$ 2.76} &\\
     $\gamma$ [km\,s$^{-1}$]&& \multicolumn{2}{c}{\textcolor{white}{4}296 $\pm$ 7} &\\
     \multicolumn{2}{l}{RMSE RV curve$^{\text{a}}$ [km\,s$^{-1}$]} & \multicolumn{2}{c}{0.22}&\\
     \hline
     \multirow{2}{*}{Parameter }& \multicolumn{4}{c}{Component}\\ 
     & \multicolumn{2}{c}{\textcolor{white}{blabl}B}& \multicolumn{2}{c}{\textcolor{white}{blabl}C}\\
     $M \sin^3 i$ [$M_\odot$] & 27 & $\pm$ 3& 11& $\pm$ 1\\
     $M$ [$M_\odot$] & 36 & $\pm$ 5& 15& $\pm$ 2\\
     $\Teff$ [kK] & 42.5 & (fixed)& 38.0& $\pm$ 1.9\\
     $R$ [$R_\odot$] & 9.37 & $\pm$ 1.77& 6.68& $\pm$ 1.74\\
     $R_{RL}$ [$R_\odot$] & \multicolumn{2}{c}{9.62} & \multicolumn{2}{c}{\textcolor{white}{1.5}6.37}\\
     $\log g$ [cm s$^{-2}$] & 4.1 & $\pm$ 0.2& 4.0& $\pm$ 0.3\\
     $\log L/L_{\odot}$ & 5.41 & $\pm$ 0.22& 4.90& $\pm$ 0.33\\
     $\Omega$ & 2.76 & $\pm$ 0.24& 2.65& $\pm$ 0.57\\
    $\Omega(L_1)$ & \multicolumn{2}{c}{2.70} & \multicolumn{2}{c}{\textcolor{white}{1.5}2.70}\\
     $M_{\text{bol}}$ & $-8.69$ & $^{+0.34}_{-0.39}$& $-7.52$& $^{+0.58}_{-1.27}$\\
     $f_{\text{C}}/f_{\text{B}}$ (I-band) & & \multicolumn{2}{c}{0.43 $\pm$ 0.03}&\\
     \hline
     \end{tabu}
    \flushleft
     \begin{tablenotes}
      \small
      \item \textbf{Notes.} $^{\text{(a)}}$ Root-Mean-Square error of the RV curve.
     \end{tablenotes}
\label{table_orbital_param_Os}
\end{table}

\begin{table*}
\renewcommand{\arraystretch}{1.1}
\centering
\caption{Stellar parameters derived from a spectroscopic and SED analysis of the four detected components in the spectra of \BAT.}
\begin{tabu}{ lr@{\hskip 0.03in}llr@{\hskip 0.03in}llr@{\hskip 0.03in}llr@{\hskip 0.03in}l }
    \hline
     \hline
     \multirow{2}{*}{Parameter}& \multicolumn{11}{c}{Component}\\
     & \multicolumn{2}{c}{A}&& \multicolumn{2}{c}{ B}&&\multicolumn{2}{c}{C}&&\multicolumn{2}{c}{D}\\
    \hline
     Spectral type &\multicolumn{2}{c}{WN2.5-3}&& \multicolumn{2}{c}{ O4\,V }&&\multicolumn{2}{c}{O6.5\,V }&&\multicolumn{2}{c}{B1\,V}\\ 
     $M$ [$M_\odot$] & 17&$\pm$ 1 && 36&$\pm$ \text{5}$^{\text{a}}$ &&15&$\pm$ 2$^{\text{a}}$ && \multicolumn{2}{c}{-}\\
     $R$ [$R_\odot$] & $2.5$&$_{-0.6}$ $^{\text{b}}$ && 9.26 &$\pm$ 1.15 && 6.31 &$\pm$ 1.11&& \multicolumn{2}{c}{-}\\ 
     $T_{\text{eff}}$ [kK] & $89$&$_{-10}$ $^{\text{b}}$ &&42.5 &$\pm$ 1.3 &&37.5 &$\pm$ 1.3&& 25.0 &$\pm$ 2.5\\
     $\log L/L_{\odot}$ (SED)$^{\text{c}}$& $5.56$&$\pm$ 0.06& & $5.37$&$\pm$ 0.13&& $4.85$&$\pm$ 0.13&& $4.23$&$\pm$ 0.09\\
     $f/f_{\text{tot}}$ (V) [\%] & $33$&$\pm$ 8&& $40$&$\pm$ 7$^{\text{d}}$ && $17$&$\pm$ 3$^{\text{d}}$ && $10$&$\pm$ 1\\
     $\varv_{\text{rot}}\sin i$ [km s$^{-1}$]&\multicolumn{2}{c}{-} &&$330$&$\pm$ 40&& $200$&$\pm$ 30 && $40$&$\pm$ 10\\
     \hline
     \end{tabu}
     \flushleft
     \begin{tablenotes}
      \small
      \item \textbf{Notes.} $^{\text{a}}$ These were derived from the orbital analysis (\Tabref{table_orbital_param_Os}). $^{\text{b}}$ The errors indicate the lower limit. $^{\text{c}}$ The errors are formal errors considering the errors on the light ratios. They are probably underestimated (see text). $^{\text{d}}$ The errors represent boundaries on the individual contributions, however only solutions for which $f_{\text{C}}/f_{\text{B}}=0.43\pm0.03$ are considered valid.
     \end{tablenotes}
\label{table_rotvs}
\label{table_orbital_analysis_Os}
\label{table_contributions}
\label{table_luminosities_SEDs}
\end{table*}


\section{Spectral analysis \& luminosities} \label{sec_spectral}

We used model spectra for the determination of the effective temperatures on the composite spectra. For the O- and B-type stars, we used non-LTE TLUSTY model atmospheres \citep{Hubeny_1998} for LMC stars. The TLUSTY code solves the radiative transfer equation in a plane-parallel geometry. The models that were used are obtained from the online catalogue of available models\footnote{\url{http://tlusty.oca.eu/Tlusty2002/tlusty-frames-BS06.html}}, which are calculated assuming a microturbulent velocity of 10\,km\,s$^{-1}$ (O-type stars) or 2\,km\,s$^{-1}$ (B-type stars). The model grids cover effective temperatures from 27.5\,kK to 55\,kK with $\Delta\Teff = 2.5$\,kK and from 15\,kK to 30\,kK with $\Delta\Teff = 1.0$\,kK for O- and B-type stars respectively, and surface gravities from $\log g = $2.5\,[cm\,s$^{-2}$] to 4.5\,[cm\,s$^{-2}$] with $\Delta\log g = 0.25$\,[cm\,s$^{-2}$].
For the WR star, we used non-LTE models computed with the PoWR \citep[Potsdam Wolf-Rayet model atmosphere code,][]{Hamann_2003} code\footnote{\url{http://www.astro.physik.uni-potsdam.de/PoWR}}, which solves the radiative transfer problem in spherical geometry in an expanding atmosphere.

\subsection{Rotational velocities}\label{sec_spectral_rotv}
In order to determine the models best representing the components B, C and D, we first determined their rotational broadening. We used a Fourier transform method based on a technique by \citet{Gray_1973}, which was applied, for example, by \citet{Simon_Diaz_2014}. For star D, we used the O\,{\sc ii}\,$\lambda\lambda$4415, 4417 line to determine the projected rotational velocity because the line is not contaminated by components B and C. We used the spectrum taken at MJD = 58437 ($\phi_{\text{1.55d}}=0.03$) since an eclipse of component B enhances the strength of spectral lines belonging to other components, namely components A and D. For components B and C, we used the He\,{\sc i}\,$\lambda$4471 line because the WR does not contaminate this line. Here, we used the spectrum taken at MJD = 58411 ($\phi_{\text{1.55d}}=0.23$) as the spectral lines of the components are at maximum separation.\nextline
The derived rotational velocities are listed in \Tabref{table_rotvs}. Their uncertainty was determined by the FWHM of the Fourier peak that indicates the rotational velocity of the component. The spectral lines of components B and C are entangled, even at the quadrature spectra. Hence, the wings of their spectral lines cannot be clearly distinguished from one another. Therefore, we rounded the rotational velocities to the nearest multiple of ten.\nextline
Accounting for the inclination of the system, we obtain deprojected rotational velocities of $360\pm60\,\text{km s}^{-1}$ and $220\pm40\,\text{km s}^{-1}$ for component B and C respectively. Assuming synchronisation, we can also derive equatorial velocities using the value of the radii obtained with PHOEBE and plugging these into the equation $\varv_{\text{rot,eq}}=2\pi R/P_{\text{rot}}$, with $P_{\text{rot}} = P_{\text{orb}} = 1.55$\,days. The derived values are $306\pm58\,\text{km s}^{-1}$ and $218\pm57\,\text{km s}^{-1}$ for component B and C, respectively. Hence, our deprojected rotational velocities agree quite well with the equatorial velocities. For all values, the errors are obtained through error propagation.

\subsection{Spectral types and effective temperatures} \label{sec_spectral_types_teff}
All derived spectral types and effective temperatures are listed in \Tabref{table_rotvs}.
\subsubsection{Component A (WN2.5-3)}
For the classification of the WR star, we used table~2 of \citet{van_der_hucht_2001}. The absence of low-ionisation lines of N, such as the N\,{\sc iii}\,$\lambda\lambda$4634, 4641 lines, and the absent/marginal N\,{\sc iv}\,$\lambda$4058 line, suggest an earlier type than WN4. The N\,{\sc v} lines are clearly present. The WR star is thus classified as $\text{WN2.5-3}$ type, in agreement with the previously determined WN3 type by \citet{Shenar_2019}. We used their derived wind parameters ($\log \dot{M} = -4.6\pm0.2\,[\MSun/$yr], $v_{\infty} = 2200\pm200$\,km\,$s^{-1}$) and abundance parameters (surface hydrogen fraction $X_\text{H} = 0-0.1\%$, surface nitrogen fraction $X_\text{N} = 0.8\%$) to create a customised PoWR model for the WR star. We adopted the effective temperature and its lower limit as derived by \citet{Shenar_2019}, which is $89_{-10}$\,kK. We determined the radius of the star from the luminosity $\log(L/L_{\odot}) = 5.56\pm0.06$ (\Secref{sec_spectral_ana_luminosities}) through Stefan-Boltzmann's law and obtain $R = 2.5_{-0.6}R_{\odot}$, where we only obtain a lower limit because of the effective temperature. Since the WR star is a naked helium core, we further used the mass-luminosity relation for a homogeneous He-star \citep{grafener_2011} and obtain a current spectroscopic mass of $M_{\text{spec}} = 17\pm1 \MSun$, where the error was obtained from a simple error propagation of the error on the luminosity.

\subsubsection{Components B and C (O4\,V and O6.5\,V)}
In the O star domain, the effective temperature correlates with the relative line strength of the He\,{\sc i}\,$\lambda$4471 and He\,{\sc ii}\,$\lambda$4542 lines. 
We therefore used the criteria from \citet{Conti_1988}, based on the equivalent width ratio of the He\,{\sc i}\,$\lambda$4471 and He\,{\sc ii}\,$\lambda$4542 lines and we estimated the spectral types of the two stars by comparing their equivalent widths from the (gaussian) fits obtained from the RV determination.
We assign a classification of O4$^{+1}_{-1}$\,V and O6.5$^{+1}_{-1}$\,V respectively for the hotter (star B) and cooler (star C) component.

We further used non-LTE TLUSTY models for the determination of the effective temperature of components B and C. Given that the two O-type stars form such a tight (possibly contact) binary, the possibility of any of them being a supergiant is excluded. Furthermore, changing the effective temperatures of the components does not significantly alter the surface gravity $\log g$ obtained from the fitting of the LC and RV curve, confirming the main-sequence character of the two stars (\Secref{sec_orbital_Os}). In the precalculated TLUSTY model grid, the value corresponding closest to the ones obtained with PHOEBE is $\log g = 4.0$\,[cm\,s$^{-2}$] for both stars. Hence, we used models with this $\log g$ value.
Finally, we accounted for the previously obtained estimates for the rotational broadening (\Secref{sec_spectral_rotv}).
For the effective temperature, we used the quadrature spectra (MJD = 58411 and MJD = 58446), yielding $\Teffo = 42\,500\pm1\,250$\,K and $\Tefft = 37\,500\pm1\,250$\,K for component B and C respectively, where the uncertainties correspond to half a grid spacing.

\begin{figure}
    \centering
    \begin{subfigure}{1\linewidth}
    \includegraphics[width = \textwidth]{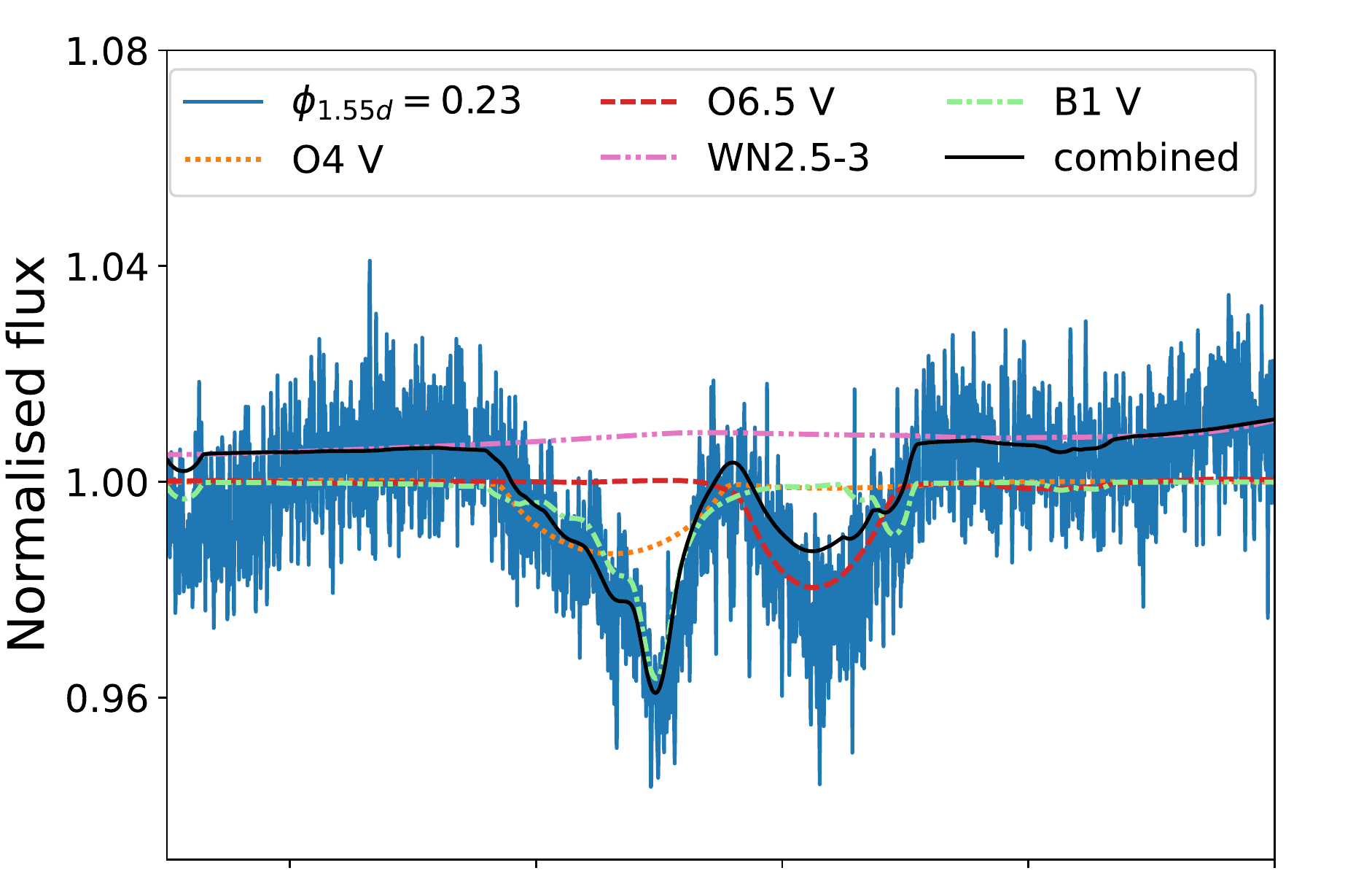}
    \end{subfigure}
    \begin{subfigure}{1\linewidth}
    \includegraphics[width = \textwidth]{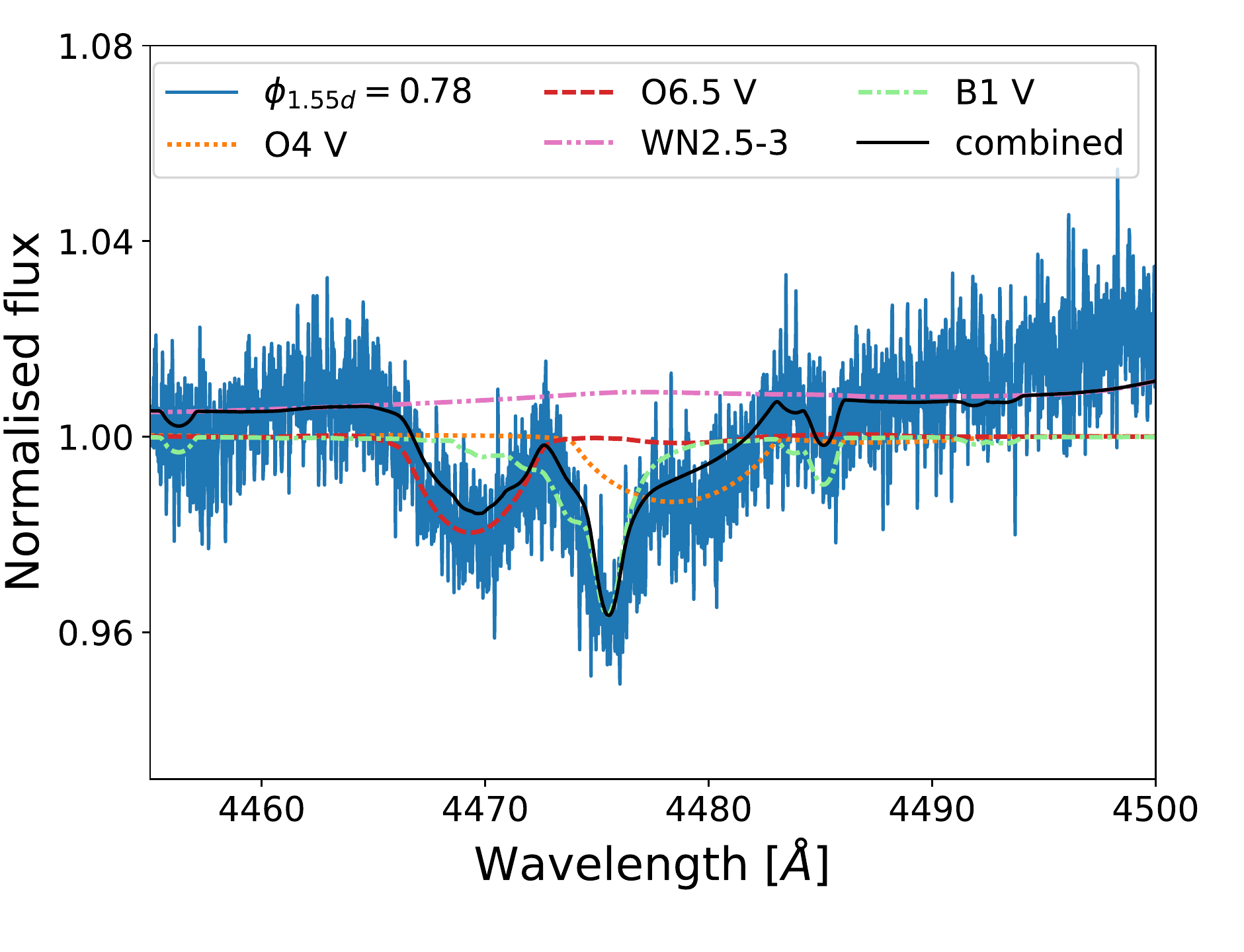}
    \end{subfigure}
    \caption{Model fits to  the He\,{\sc i}\,$\lambda$4471 region at quadrature phases in the 1.55-d orbit. The solid-black curve shows the combined model. Top panel: $\phi_{1.55d} = 0.23$. Bottom panel: for the spectrum corresponding to a phase $\phi_{1.55d} = 0.78$.}
    \label{figure_contributions_HeI}
\end{figure}

\begin{figure*}
    \centering
    \includegraphics[width = \textwidth]{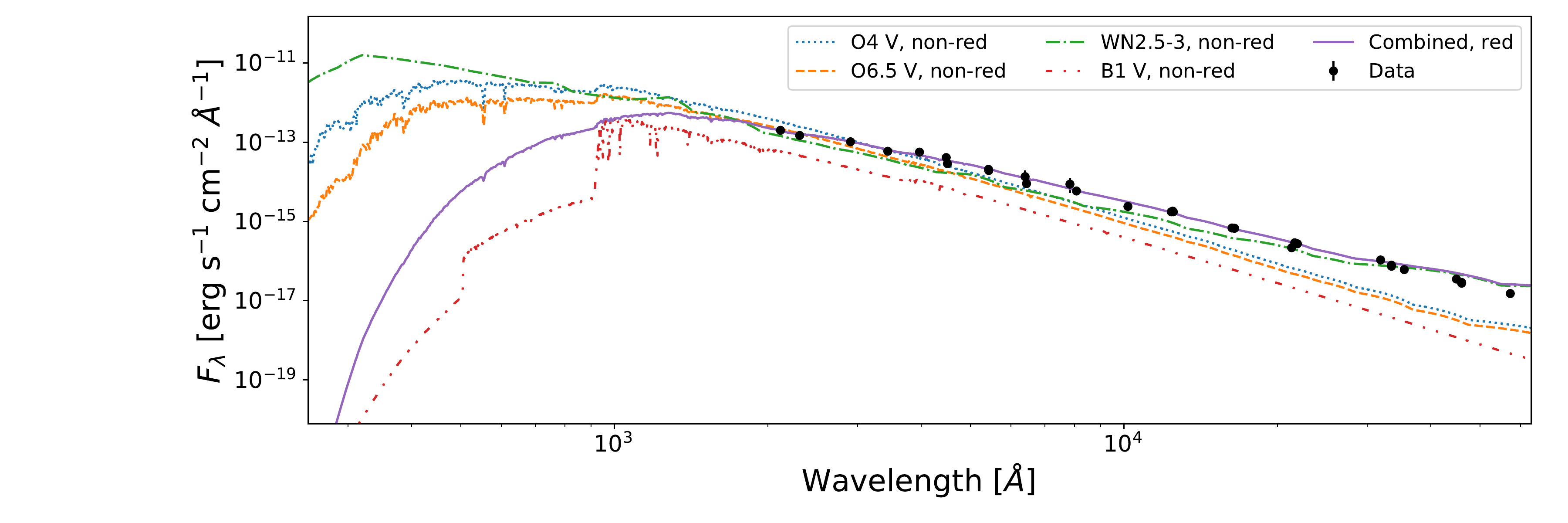}
    \caption{Model SED of the system fitted to the photometric data points. The combined model SED is reddened, whereas the individual SEDs are not.}
    \label{figure_SEDs}
\end{figure*}
\subsubsection{Component D (B1\,V)}
The presence of Si\,{\sc iii} and O\,{\sc ii} lines combined with the absence of He\,{\sc ii} lines indicate that component D is a cooler B-type massive star. Since the B-type star is not dominating the composite spectrum, the possibility that this star is a supergiant is excluded and hence we fix $\log g = 4.0$\,[cm\,s$^{-2}$]. For its classification, we used the criteria presented in table 2 of \citet[][]{Evans_2015}. The spectral type and effective temperature were derived from the spectrum at MJD = 58437. It corresponds to $\phi_{\text{1.55d}}=0.03$ and the eclipse of one of the O-type stars makes the spectral lines of the WR star and B-type star more prominent.\nextline
The Si\,{\sc iii}\,$\lambda$4553 and Mg\,{\sc ii}\,$\lambda$4481 lines are clearly present, however the Si\,{\sc iv}\,$\lambda$4089 line of the B-type component is very weakly present. Furthermore, we see that Si\,{\sc iii}\,$\lambda$4553 $\gtrsim$ Mg\,{\sc ii}\,$\lambda$4481 and classify component D as a B1$\pm$1\,V star.\\
For the determination of the effective temperature we used the same technique as for the two O-type stars. The non-LTE TLUSTY models were broadened and rescaled to an estimated flux contribution of 10\% (this appears to be in agreement with \Secref{sec_spectral_contributions}). We obtain an effective temperature $\Teff = 25\,000\pm2\,500$\,K.

\subsection{Optical light contributions} \label{sec_spectral_contributions}
In order to obtain the contribution of each component to the total flux in the optical spectrum, we used the model spectra of the individual components to create a composite model spectrum for the system. As previously mentioned, we used for the WR star the customised PoWR model spectrum and for the O- and B-type stars a TLUSTY model. However, since the spectrum of the WR star is difficult to reproduce, we cannot use a fitting algorithm to determine the best fit, but instead we found the optical light contributions by a visual inspection. For this inspection, we payed careful attention to the fit of the two O-type stars and the B-type star and less to the WR star for the aforementioned reason, and hence also not to the absorption lines located in the WR emission lines. 
In particular, we focused on the He\,{\sc i}\,$\lambda$4771 line for the determination of the individual contributions. The two O-type stars and the B-type star are clearly visible, whereas the WR star does not contribute here. \nextline
The spectra on which the analysis was performed are the quadrature spectra of the 1.55-day orbit (MJD = 58411 and MJD = 58446). The original models were shifted according to their derived RV (\Secref{sec_orbital_Os}) and were scaled to different light contributions. The estimated contributions are given in \Tabref{table_contributions} while \Figref{figure_contributions_HeI} shows the fits of the He\,{\sc i}\,$\lambda$4771 line for the two quadrature spectra.\nextline
The contribution of the B-type star is well constrained. For the other three components, this is not the case. The orbital analysis provides us with a light ratio in the I-band, which is virtually the same as the visual light ratio. We use this information and only consider solutions for which $f_{\text{C}}/f_{\text{B}}=0.43\pm0.03$
.

\subsection{Luminosities} \label{sec_spectral_ana_luminosities}
We derived individual luminosities of the components by fitting models of the spectral energy distributions (SEDs) to photometric data (see \Tabref{appendix_photometric_data}). The SED models that were used correspond to those with parameters derived in \Secref{sec_spectral_types_teff}. We scaled the models to the distance of the LMC \citep[$d_{\text{LMC}} = 46.97$\,kpc,][]{Pietrzynski_2013} and reddened the models using the reddening law by \citet{Howarth_1983}. We also took into account the individual contributions of the components derived in \Secref{sec_spectral_contributions}, again only taking into consideration the solutions for which $f_{\text{C}}/f_{\text{B}}=0.43\pm0.03$. These contributions correspond to those in the $V$-band and hence the SEDs were scaled in the $V$-band to match the relative contributions. The O4\,V component was used as a reference and the other components were scaled relative to the SED of this star. By adding the individual SEDs, we created a model SED for the system, which could then be fitted to the photometric data.\nextline
The fitting of the model SED of the system was done with a $\chi^2$-minimisation procedure with the formula
\begin{equation}
    \chi^2 = \frac{1}{N-D}\sum_{i=1}^{N}\left( \frac{F_{i,\text{obs}} - F_{i,\text{mod}}}{\sigma_{i,\text{obs}}}\right)^2
\end{equation}
where $N$ is the number of data points and $D$ is the number of model parameters, which are the reddening and the relative extinction parameter. The parameters $F_{i,\text{obs}}$ and $F_{i,\text{mod}}$ are the flux of the i-th data point and the flux of the model SED at the same wavelength respectively, and $\sigma_{i,\text{obs}}$ is the uncertainty on the i-th data point. \nextline
The error on the obtained luminosities has two contributions. The first one comes from the uncertainty on the fit itself. In order to determine this part of the error, we performed 1000 Monte-Carlo simulations where each time a new data set was created from the original photometric data points using
\begin{equation}
    F_{i,\text{new}} = F_{i,\text{obs}} + \mathcal{N}(0,1) \times \sigma_{i,\text{obs}}
\end{equation}
where $\mathcal{N}(0,1)$ is a random number from the normal distribution. For each of these new data sets, we repeated the procedure described above. This procedure only returns an error of about 0.01\,dex. A much larger contribution to the error originates from the uncertainty on the light ratios. We repeated the SED-fitting procedure for the possible combinations of individual light contributions in order to obtain the final formal error, given in \Tabref{table_luminosities_SEDs}. The fit to the data points is shown in \Figref{figure_SEDs}. The obtained reddening for the system is $E_{\text{B-V}}=0.12\pm0.02$. We point out that the fainter source, probably component D, may have a smaller reddening than the brighter source consisting of high-mass components exhibiting strong mass loss.\nextline
The final luminosities and formal errors are given in \Tabref{table_luminosities_SEDs}. The addition of two more components slightly lowers the luminosity of the WR star compared to the luminosity derived for a WN3 + O6.5\,V binary \citep[$\logL = 5.78\pm0.15$,][]{Shenar_2019}. 
The luminosities for the two O-type stars are well in agreement with those obtained with PHOEBE. We note that although the formal errorbars on the SED luminosities are smaller than those obtained with PHOEBE, the latter are still more reliable due to additional uncertainties we did not take into account, such as the uncertainties on the effective temperatures of the components for the SED fitting. Also, as mentioned in the previous paragraph, it might be the case that not all components have the same reddening, whereas for the fitting it is assumed they do. For the remainder of the analysis, we used the luminosities obtained from the orbital analysis for the two O-type stars.

\section{Evolution of the close O+O binary}\label{sec_evolution}
\subsection{The age of the system}
\begin{figure}
    \centering
    \includegraphics[width = 0.5\textwidth]{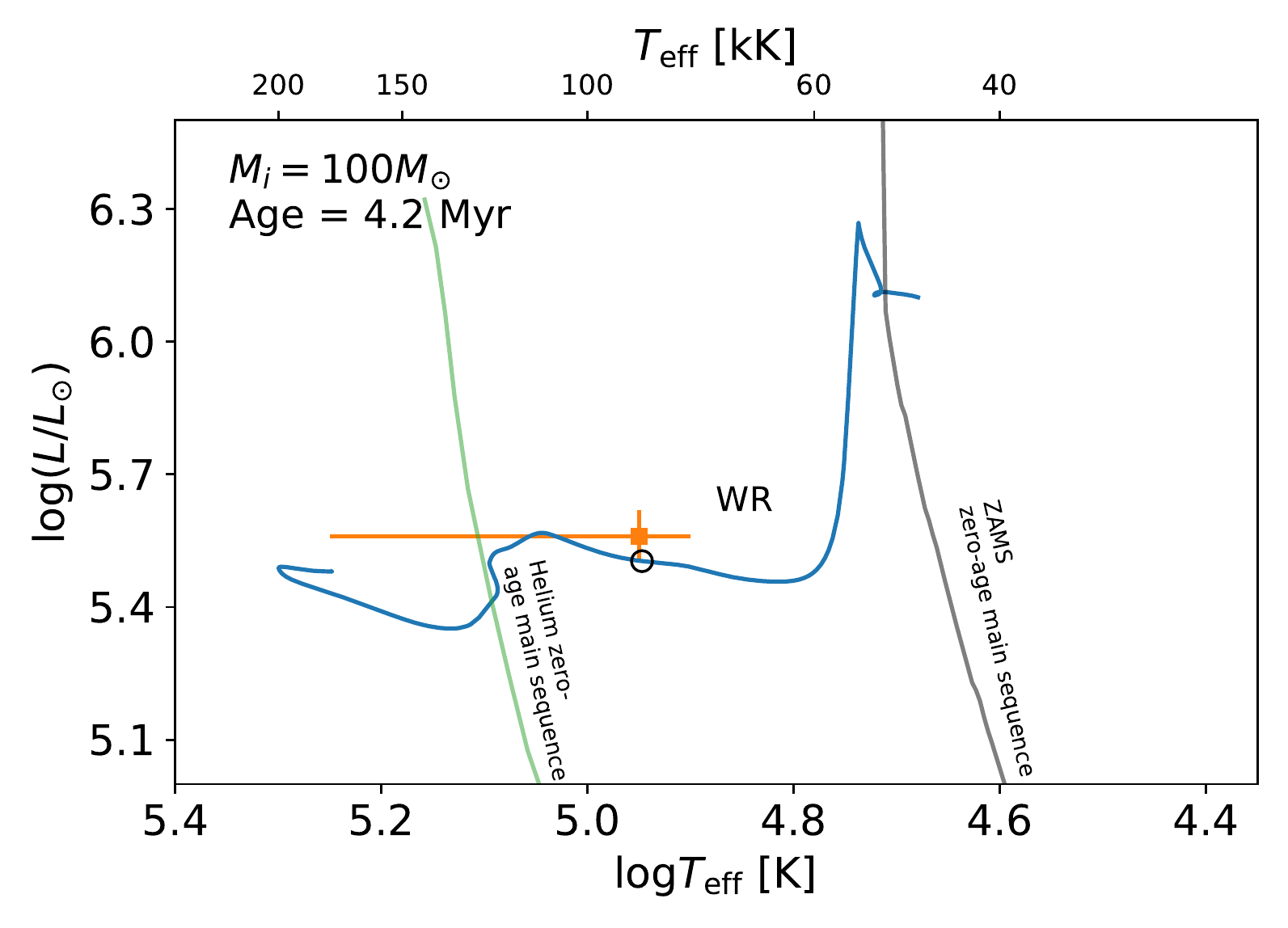}
    \caption{Evolutionary (BPASS) track of a homogeneously evolving star with $\Mini = 100\MSun$. The observed WR star is shown with an orange square. The best fitting position on the track is indicated with a black circle and corresponds to an age of 4.2\,Myr.}
    \label{figure_evolution_BPASS}
\end{figure}

Given the  proximity of the WR star and the O+O binary, we worked under the assumption that all components are coeval and we first focused on the WR star as an appropriate age estimator for the system given the short lifespan of the WR phase. In this context, we compared the parameters of the WR star of \BAT\space against evolutionary tracks from the Binary Population and Spectral Synthesis code\footnote{\url{http://bpass.auckland.ac.nz}} \citep[BPASS;][]{Eldridge_2016}. Given that we do not see a companion to the WR star, we used CHE and non-CHE tracks for single stars with LMC metallicity $Z=0.008$. In order to find the best fitting track, we used a $\chi^2$-minimisation procedure 
\begin{equation} \label{eq_chi2_evo}
    \chi^2 = \sum_{i=1}^{4} \left(\frac{O_i - E_i}{\sigma_i} \right)^2
\end{equation}
where $O_i \in \{ \Teff, \log(L/L_{\odot}), X_H, M_{\text{spec}}\}$, $E_i$ the value of the parameter at a given point on the track, and $\sigma_i$ is the error on the observed parameter. The values for $\Teff$, $\log(L/L_{\odot})$, $X_H$, and $M_{\text{spec}}$ are given in \Secref{sec_spectral_types_teff}.\nextline
Figure~\ref{figure_evolution_BPASS} shows the evolutionary track corresponding to a homogeneously evolving star with $\Mini = 100\Modot$ and is the best fitting track for the WR star. The black circle indicates the position of the WR star on the track, corresponding to an age of 4.2\,Myr. The uncertainty on the derived age is mainly dominated by model uncertainties, which is why we assign an error of 0.5\,Myr. The best fitting track for a non-CHE star was that of a star with $\Mini = 40\Modot$ and returned an age of 5.3\,Myr. However, the returned surface hydrogen fraction fell outside of the accepted range, which is why we adopt the CHE track in this exercise. Most results presented in the next sections do not depend on the age estimate of the WR star, however it is used as a comparison to the obtained age for the O+O system.

\subsection{The O-type components as single stars}
We first confronted the obtained constraints of the O stars to single star evolution models.
In this purpose, we used the baysian tool BONNSAI\footnote{\url{www.astro.uni-bonn.de/stars/bonnsai}} \citep[BONN Stellar Astrophysics Interface;][]{Schneider_2014} to retrieve the most probable initial mass, age, current mass, effective temperature and luminosity of the O stars in \BAT. It uses Bayesian statistics to compare the observables to stellar models from \citet{Brott_2011} whilst taking into account priors, such as initial mass functions and distributions of stellar rotational velocities. The parameters returned by BONNSAI are given in \Tabref{table_evolution_BONNSAI} under `Single'. \nextline
The BONNSAI results indicate that there is an age discrepancy between the two components. With an age of 1.9\,Myr, the primary appears to be much younger than the 6.8\,Myr-old secondary, which is not expected given the proximity of the two components. In addition, the posterior luminosity and effective temperature of the secondary are much lower than those observed. In order to have a single star with such a high luminosity and effective temperature, we would need a more massive single star. Hence, the secondary is overluminous. These findings suggest that the O stars in \BAT\ did not evolve as single stars. Indeed, the age discrepancy is a sign of rejuvenation of the primary component \citep[e.g.][]{Schneider_2014_rejuv, Mahy_2020} while the overluminosity and higher effective temperature of the secondary star are indicative of the secondary being the initially more massive and now more evolved object of the pair. This is typical of the Algol paradox \citep{Kuiper_1941} and again suggest that there has been prior interaction between the two components.

\begin{table*}[]
    \renewcommand{\arraystretch}{1.1}
    \centering
    \caption{Parameters for the two O-type stars. The relevant observed parameters for an evolutionary analysis are shown. Values for single star evolution are obtained with BONNSAI. Values for the binary evolution are those of the MESA simulation with conservative mass transfer that best fits the values of the observed parameters.}
    \begin{tabular}{l|r@{\hskip 0.03in}lr@{\hskip 0.03in}l|r@{\hskip 0.03in}lr@{\hskip 0.03in}l|cc}
    \hline
     \hline
        \multirow{3}{*}{Param.}& \multicolumn{4}{c}{Observed} & \multicolumn{4}{c}{Single} & \multicolumn{2}{c}{Binary}
        \\
        & \multicolumn{4}{c}{}&  \multicolumn{4}{c}{}& \multicolumn{2}{c}{(conservative)}\\
         & \multicolumn{2}{c}{O$_1$} &  \multicolumn{2}{c}{O$_2$}& \multicolumn{2}{c}{O$_1$} &  \multicolumn{2}{c}{O$_2$} & \multicolumn{1}{c}{O$_1$} &  \multicolumn{1}{c}{O$_2$}\\
        \hline
        $\Mini$ [$M_{\odot}$] && &&& $36$&$\pm3$ & $15$&$\pm2$& 23&29 \\
        Age [Myr] & $4.2$&$\pm0.5$ $^{\text{a}}$&$4.2$&$\pm0.5$ $^{\text{a}}$&$1.9$&$^{+0.5}_{-1.0}$& $6.8$&$^{+2.8}_{-3.5}$ &4.6 & 4.6\\
        $M_{\text{cur}}$ [$M_{\odot}$] &$36$&$\pm5$ &$15$&$\pm2$& $36$&$\pm3$ & $15$ &$\pm2$ & 32 & 19\\
        $R_{\text{cur}}$ [$R_{\odot}$] &$9.37$&$\pm1.77$&$6.68$&$\pm1.74$& $7.86$&$^{+1.67}_{-0.42}$ &$5.99$&$^{+1.41}_{-1.01}$&9.2&7.2\\
        $T_{\text{eff,cur}}$ [kK] & $42.5$&$\pm1.3$ &$37.5$&$\pm1.3$& $42.6$&$\pm1.1$& $30.7$&$^{+2.2}_{-2.8}$& 40.3 & 36.3\\
        $\log(L_{\text{cur}}/L_{\odot})$& $5.41$&$\pm0.22$&$4.90$&$\pm0.33$&$5.33$&$^{+0.12}_{-0.09}$& $4.50$&$^{+0.14}_{-0.17}$ &5.30& 4.91\\\hline
        $P_{\text{initial}}$ [days] & \multicolumn{4}{c}{} &\multicolumn{4}{c}{}  & \multicolumn{2}{c}{1.3} \\
        $P_{\text{cur}}$ [days] & \multicolumn{4}{c}{1.55} & \multicolumn{4}{c}{} & \multicolumn{2}{c}{1.6} \\
        \hline
    \end{tabular}
    \flushleft
        \begin{tablenotes}
      \small
      \item $^{\text{(a)}}$ From the age estimate of the WR star
    \end{tablenotes}

    \label{table_evolution_BONNSAI}
\end{table*}

\subsection{The O+O system as an interacting binary}
We now consider the currently less massive star (component C) to be the initially more massive star that has transferred a significant amount of its mass to its companion, so that the accretor is the currently the more massive star. We then performed several evolutionary simulations with varying initial conditions (masses, period, rotational mixing efficiency and conservativeness of mass transfer) in order to find initial conditions that reproduce the observed properties of the O+O binary. A simulation is said to reproduce the system when the parameters returned by the simulation match the observed ones within their 1-$\sigma$ error. However, since we were working with a coarse grid size ($\Delta P = 0.1$\,days, $\Delta M = 2\MSun$ and for non-conservative mass transfer $\beta \in \{ 0.1, 0.2, 0.5, 1\}$), we allowed for a 10\% discrepancy between the period in the simulation and the observed period that has errors much smaller than 10\%.

\subsubsection{MESA and its setup}
We performed binary simulations with the stellar evolution code MESA\footnote{All input files to reproduce the simulation will be made public after acceptance of the manuscript.} \citep{Paxton_2011, Paxton_2013, Paxton_2015, Paxton_2018, Paxton_2019}, release 12115 and SDK version 20190830, which is able to simulate both conservative and non-conservative mass transfer. We used the setup for rotating massive binary stars of \citet{Marchant_2017}, adjusted to use custom opacity tables from the OPAL project, computed for a composition representative of the LMC as given in \citet{Brott_2011}\footnote{\url{https://github.com/orlox/mesa\_input\_data/tree/master/2016\_binary\_models}} \citep{Iglesias_1996}. \nextline
We performed a grid of simulations with different initial periods, initial masses and the type of mass transfer (conservative vs. non-conservative) to determine the best fitting evolutionary scenario. We also varied the rotational mixing efficiency parameter $f_m$. Rotational mixing is determined from theoretical equations such as the diffusion equation. In MESA, it is possible to multiply this theoretical factor with a variable 
in order to determine the rotational mixing efficiency. To consider potential uncertainties on the strength of rotational mixing, we var\textbf{ied} this value from its standard value of 0.033 \citep{Heger_2000} up to 20 times this value.\nextline
We account\textbf{ed} for mass transfer using the model presented by \citet{Kolb_1990}, hereafter called the Kolb model. The Kolb model predicts a non-zero amount of mass transfer from the extended atmosphere of a star, even if its photospheric radius is smaller than its Roche lobe radius. The standard Roche-lobe model only initiates Roche-lobe overflow at the time the star is actually filling its Roche-lobe. This produces a sudden turn-on of mass loss. The Kolb model does not have this sudden turn-on but rather has a continuous increase of mass loss, which makes the Kolb model more numerically stable than the standard model.\nextline
For non-conservative mass transfer, the mass transfer efficiency is determined by $\epsilon = 1 - \alpha - \beta - \delta$, where $\alpha$, $\beta$, and $\delta$ correspond to the mass lost through wind of the donor star, wind of the accreting star, and a disk, respectively \citep{Tauris_2006}. In the simulations where we model non-conservative mass transfer, we investigated the simple case of non-conservative mass transfer where part of the accreted mass is radiated from the surface of the accretor. We thus varied the $\beta$-parameter whilst keeping the others at their standard value of zero. 

\subsubsection{Chemically homogeneous evolution (CHE)}
\begin{figure}
    \centering
    \includegraphics[width = 0.5\textwidth]{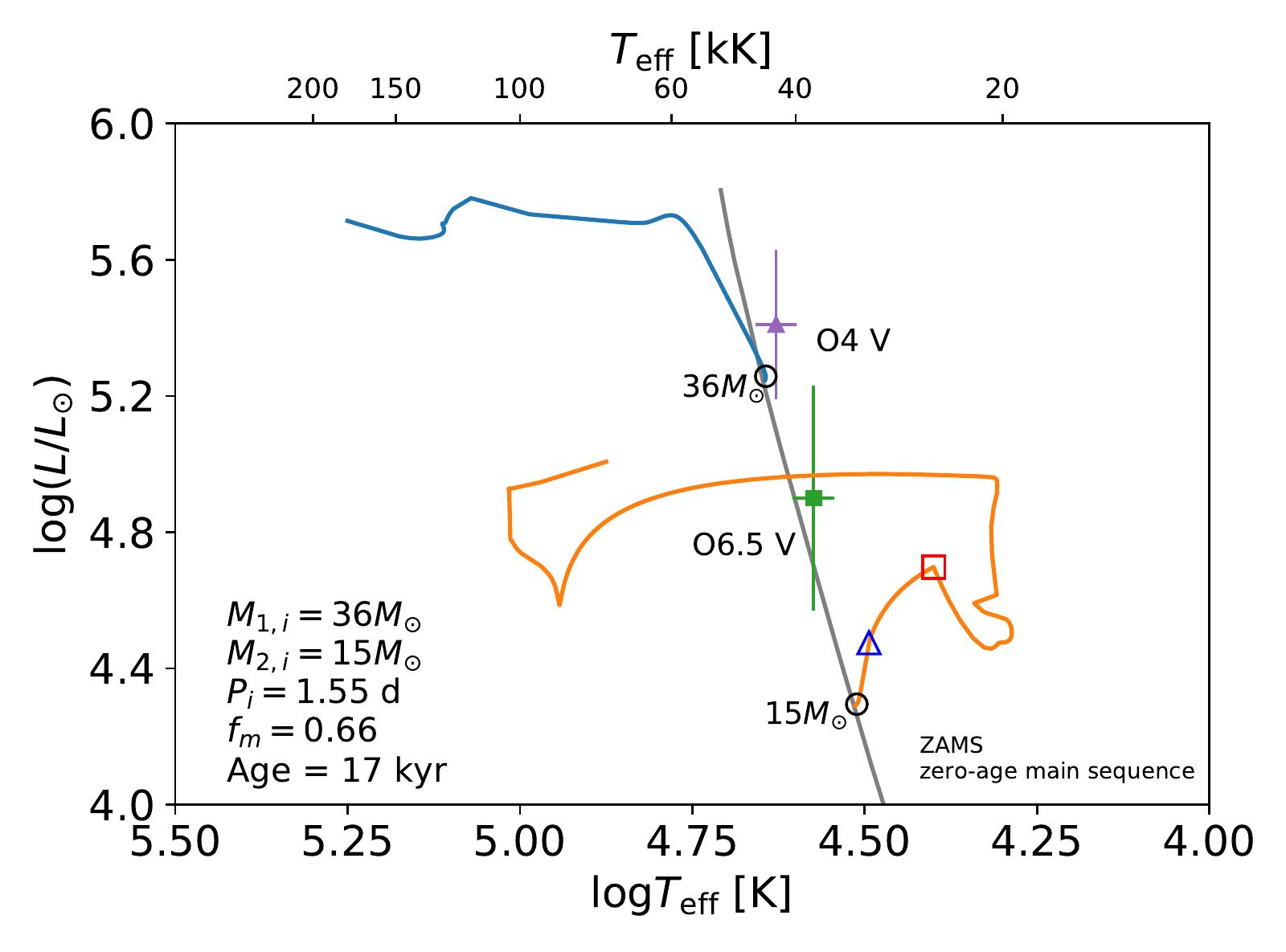}
    \caption{Binary evolutionary tracks for two stars with a rotational mixing efficiency $f_m = 0.66$ and with initial masses $M_{1,i} = 36\Modot$ and $M_{2,i} = 15\Modot$, and an initial period of 1.55\,days. The best fitting positions on the track are indicated with black open circles and correspond to an age of 17\,kyr. The $36\Modot$ star collapses into a BH at an age of 7.8\,Myr, indicated also on the $15\Modot$ track with a blue triangle. After that, the $15\Modot$ star evolves further until it starts transferring mass to the BH (red square). Eventually, the $15\Modot$ star will also collapse into a BH (the track stops at carbon depletion).}
    \label{figure_evolution_hom}
\end{figure}

We first investigated the possibility of CHE for one or both of the components. Since homogeneous evolution may prevent binary interaction, we assumed an initial system with identical masses and the same period as we observe for the system now. We performed simulations with a rotational mixing efficiency $f_m \in \{0.033, 0.066, 0.099, 0.33, 0.66 \}$, where the last value is 20 times the standard value \citep{Heger_2000}. All simulations with $f_m < 0.66$ result in a merger of the two stars around an age of 2\,Myr. The simulation with $f_m = 0.66$ results in a homogeneously evolving primary, while the secondary does not evolve chemically homogeneous. Figure~\ref{figure_evolution_hom} shows the resulting tracks for the latter simulation. We performed the same $\chi^2$-minimisation procedure as presented in \Eqref{eq_chi2_evo}, but now $O_i \in \{ \Teffo, \Tefft, L_1, L_2, R_1, R_2, M_1, M_2, P \}$. The resulting best-fit corresponds to an age of 17\,kyr for the components. This age is far from the benchmark age estimate obtained from the WR star. Moreover, the radii of both components ($R_1 = 6.7\,R_{\odot}$ and $R_2 = 4.3\,R_{\odot}$) and the luminosity of the secondary ($\log(L_2/L_{\odot}) = 4.3$) are only just within 2$\sigma$ of the observed values, and the effective temperature of the secondary ($T_{\text{eff,2}} = 33.9$\,kK) is at this point just within 3$\sigma$ of the observed value.
Hence, we do not consider a chemically homogeneous evolutionary scenario as representative of the evolution of the O+O binary.

\subsubsection{Mass transfer}
\begin{figure}
    \centering
    \includegraphics[width = 0.5\textwidth]{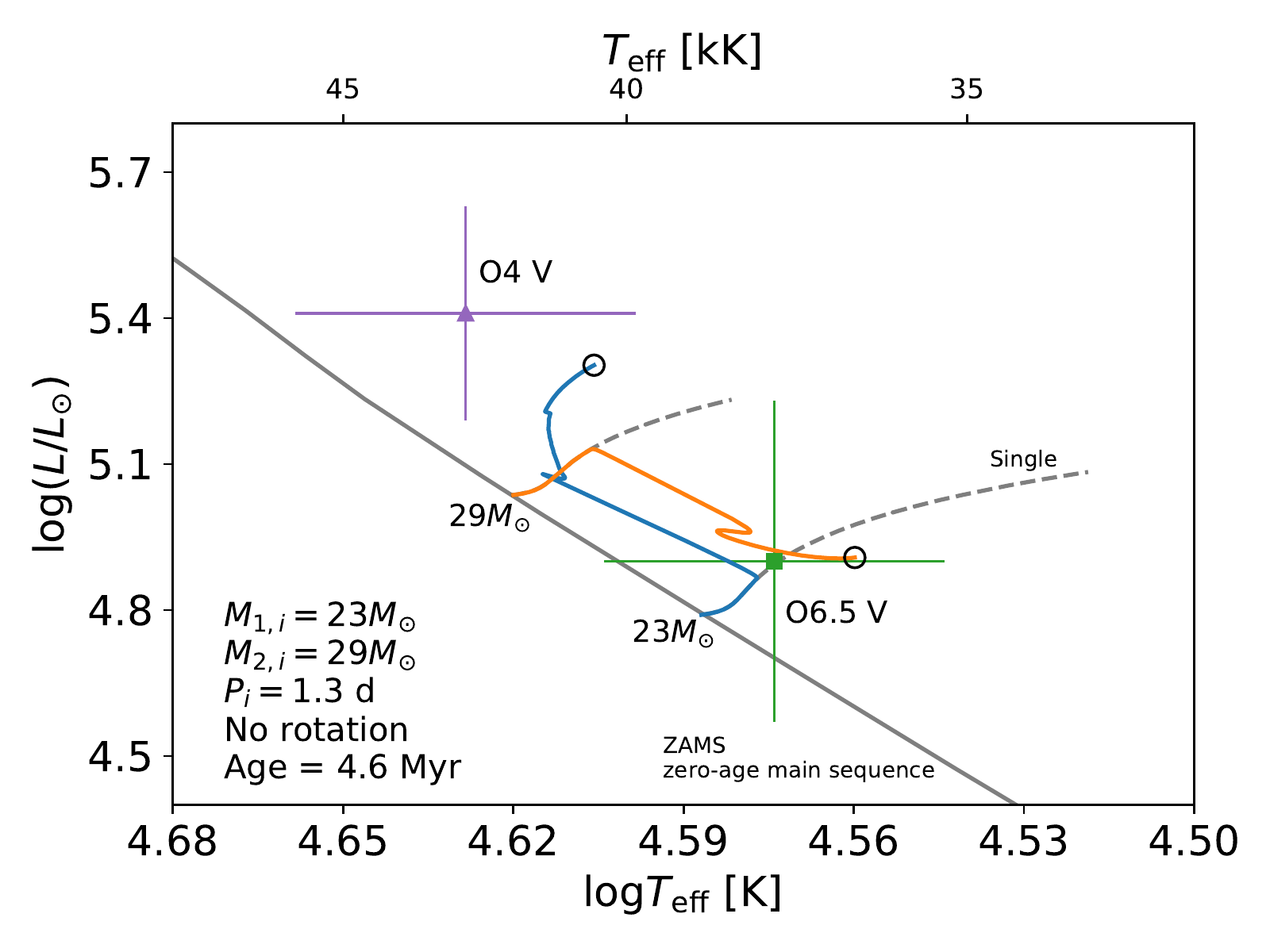}
    \caption{Binary evolutionary tracks for two non-rotating stars with initial masses $M_{1,i} = 23\Modot$ and $M_{2,i} = 29\Modot$, and an initial period of 1.3\,days. The best fitting positions on the track are indicated with black open circles and correspond to an age of 4.6\,Myr.}
    \label{figure_evolution_conservative_O}
\end{figure}

We focused here on determining the initial conditions of the close O+O binary while varying the initial masses and period together with the type of mass transfer (conservative/non-conservative). We have to take into account that for non-conservative mass transfer, the total initial mass of the O+O binary is much larger than the total mass of the observed system due to the mass lost in the mass transfer process. For the simulations with conservative mass transfer, 
in order to account for mass lost through stellar winds ($\sim10^{-6}\Modot$/yr), we used an initial total mass that is $1\Modot$ larger than the one observed ($52\Modot$). Since we have shown that chemically homogeneous models do not reproduce the system, we excluded rotation from these simulations. \nextline
We first performed simulations over a broad parameter space with an initially less massive star which is a point mass. This means that instead of an evolving star, the accretor is now a point accreting mass which does not evolve. Hence, it allowed us to explore a broader parameter space at a smaller computational cost. We then examined which of the simulations have a donor star that agrees with the observations of the observed O4\,V star (the primary). For simulations with conservative mass transfer, we need an initial mass ratio -- the initially less massive star divided by the initially more massive star -- closer to one and shorter periods, whereas for non-conservative mass transfer we need more extreme initial mass ratios and longer periods. \nextline
Having constrained the range of initial parameters that can reproduce the properties of the primary, we then performed simulations in this reduced parameter space but including the evolution of both stars. For simulations with non-conservative mass transfer, the system evolves into an over-contact binary before any of the observed parameters are returned by the simulation. At this point, the simulation stops and most likely the outcome of such a system is the merger of the two stars. For completely non-conservative mass transfer ($\beta = 1$), the final mass ratio is close to unity before the system became an over-contact system, showing us that conservative mass transfer is needed in order to reproduce the mass ratio that is observed.\nextline
Figure~\ref{figure_evolution_conservative_O} shows the best-fitting evolutionary tracks for the two O-type stars. As our simulations do not account for reverse mass transfer, the tracks stop when the accretor fills its Roche lobe. The initial system has initial masses of $29\MSun$ and $23\MSun$ for the donor and accretor respectively, and an initial period of 1.3\,days. 
The best-fitting point corresponds to an age of 4.6\,Myr, in excellent agreement with the age of the WR star. The resulting values are given in \Tabref{table_evolution_BONNSAI}. We see that all parameters agree well with the observations within their allowed error.\nextline
The simulation presented here is not the only simulation which returns a system that is able to reproduce the observed system. Over all simulations, those that are able to reproduce the system have initial periods between 1.0\,days and 1.5\,days with an initial mass ratio between 0.7 and 0.95. We conclude that a phase of conservative mass transfer between the two components is the most likely evolutionary scenario of the close O+O binary of \BAT. For all simulations performed, the accretor also fills its Roche lobe around an age of 3.5-4.5\,Myr, resulting in the formation of a (over)contact system. Within the uncertainties on the parameters derived by the LC, we cannot exclude that the O+O binary is currently in (over)contact. Such a system is believed to eventually merge before the two components collapse into BHs. This indicates that the close O+O binary does not seem to be a progenitor of a BH merger or ULX according to the scenario presented by \citet{de_Mink_2016} and \citet{Marchant_2016, Marchant_2017}.
However, it remains possible that the O+O system, after  merging into a single O star, may evolve towards a red supergiant (RSG). The WR star will, by that time, already have collapsed into a BH. If the WR star is now bound to the O+O system, and hence the BH later to the RSG, they may enter a phase of common envelope, resulting in the merger of the two objects, or the ejection of the envelope of the RSG. 

\section{Summary}\label{sec_summary}

Using new high-resolution spectroscopic data, we conclude that BAT99\,126 is at least a quadruple system. The exact configuration of the system remains unclear. While \BAT\space could be a hierarchical system with four or five components, it could also be a combination of two SB2 systems (see \Figref{fig_scheme}). The four well-identified components consist of two O-type stars (O4\,V and O6.5\,V), a WR star (WN2.5-3), and a B1\,V star. 
The O4\,V and O6.5\,V star are paired in a 1.55d-period binary and  have masses of $M_1 = 36\pm5\Modot$ and $M_2 =
15\pm2\Modot$, respectively. The orbital and stellar parameters of the two O-type stars can be found in \Tabref{table_orbital_param_Os} and the stellar parameters of the other components can be found in \Tabref{table_orbital_analysis_Os}.

The previously suggested orbit of 25.5 days for the WR star \citep{Foelmi_2003b} is excluded as no significant motion is detected over the two-month period covered by our data.
The B-type star displays clear RV variability (see also \Figref{figure_motion_B}), however more observations are needed to constrain a putative orbit. The data is also inconclusive on whether the WR star and B-type star are in orbit with each other. Other possibilities for the B-type star are that it orbits the close O+O binary or that it has another, so far undetected, companion. This latter hypothesis is supported by the flux ratio analysis of the two sources seen in an archival HST-UV image.
\nextline
The age of BAT99\,126 is estimated from the WR star to be $4.2\pm0.5$\,Myr, as derived from BPASS evolutionary tracks for homogeneously evolving stars. Combining the estimated age with the masses and orbital separation, we performed a tailored modelling of the evolution of the O+O binary. We exclude the possibility that the two O-type components have evolved as single stars as we detect clear signs of the rejuvenation of the primary star, which suggests that the currently least massive star was originally the most massive star. 
We also show that the O+O binary is not well reproduced by chemically homogeneous models. \nextline
We also performed binary evolution computations including both conservative and non-conservative mass transfer. We found that simulations with conservative mass transfer reproduce the observed system best, with initial periods in between 1\,day and 1.5\,days, initial mass ratios between 0.7 and 0.95, an initial mass for the initially more massive star between $27\Modot$ and $31\Modot$, and a total initial mass around $52\Modot$. The simulations indicate that the two components are overfilling their Roche lobe at around an age of 4\,Myr, suggesting that the binary is currently an (over)contact system. The uncertainties on the LC fit do not exclude this scenario.\nextline
The computations resulted in the formation of an (over)contact binary, suggesting that the O+O binary will eventually merge before any of the O stars collapse into BHs. This indicates that the close O+O binary of \BAT\space is likely not the progenitor of a BH merger or ULX. Likely, an even more massive O+O binary would be required to evolve homogeneously into a BH merger or ULX according to the scenario suggested by \citet{de_Mink_2016} and \citet{Marchant_2016, Marchant_2017}. 
\nextline
These uncertainties aside and with an initial total mass well over 150~$M_\odot$, \BAT\ is one of the most massive high-order multiple systems known to date. Longer-term observations are required to shed light on the exact structure of this complex system, specifically elucidating the remaining hierarchical configuration of the WR and B-type components with respect to one another and with respect to the O+O system.

\begin{acknowledgements}
      Based on observations obtained with the UVES spectrograph at the European Southern Observatory under programme ID 0102.D-0050(A). The authors acknowledge support from the European Research Council (ERC) innovation programme of the Horizon 2020 (programme DLV-772225-MULTIPLES), the FWO Odysseus programme under project G0F8H6N, the FWO junior postdoctoral fellowship under project 12ZY520N and the FWO PhD fellowship under project 11E1721N. L.M. thanks the European Space Agency (ESA) and the Belgian Federal Science Policy office (BELSPO) for their support in the framework of the PRODEX Programme.
\end{acknowledgements}

%
%

\bibliography{References}

\newpage
\begin{appendix}
\section{Motion of the B-type star}

\begin{figure}[h!]
    \centering
    \includegraphics[width = 0.5\textwidth]{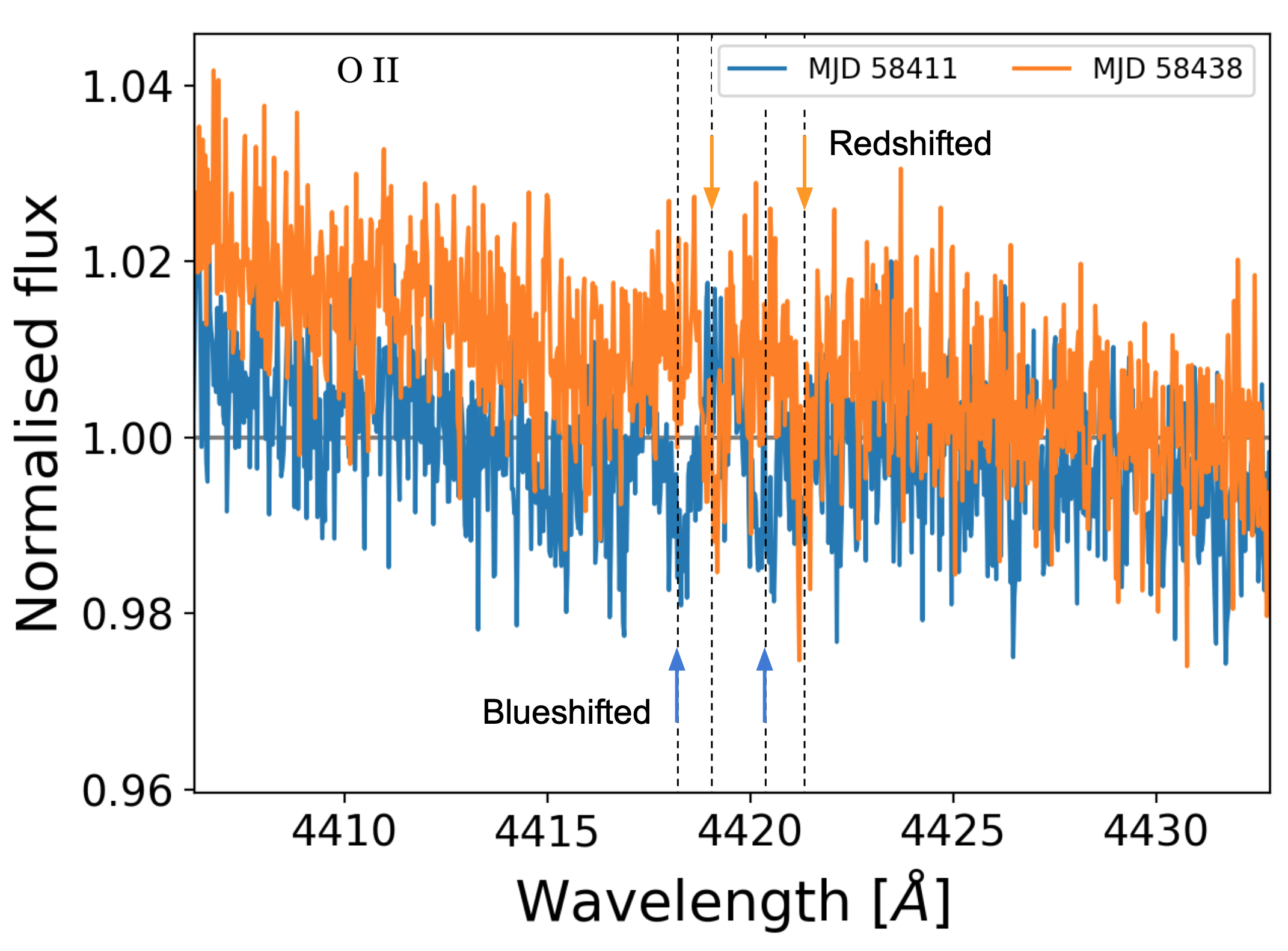}
    \caption{The O\,{\sc ii}\,$\lambda\lambda$4415\textbf{, }4417 line of the B-type star (component D) in the spectra taken at MJD = 58411 and MJD = 58438. Arrows indicate the blue-shifted and red-shifted motion between the two spectra.}
    \label{figure_motion_B}
\end{figure}

\section{Photometric data}\label{appendix_photometric_data}
\begin{table*}
\renewcommand{\arraystretch}{1.1}
    \centering
    \caption{Photometric observations of BAT99126. Data obtained from VizieR (\url{http://vizier.unistra.fr/vizier/sed/}).}
     \begin{tabular}{lll@{\hskip 0.02in}lll}
        \hline \hline
        Passband & Effective wavelength [\AA]& \multicolumn{2}{l}{Flux [erg s$^{-1}$cm$^{-2}$\AA$^{-1}$]}& Magnitude &Ref.\\
        \hline
       XMM-OT:UVW2 &\textcolor{white}{00}2~118.97& 1.99e-13 &$\pm$ 6.68e-16 & 10.66& a \\
       XMM-OT:UVM2 &\textcolor{white}{00}2~311.07& 1.48e-13 &$\pm$ 5.61e-16 & 10.99 & a\\
       XMM-OT:UVW1 &\textcolor{white}{00}2~908.07 &1.02e-13 &$\pm$ 1.44e-17 & 11.39 & a \\
        XMM-OT:U &\textcolor{white}{00}3~441.18 &5.90e-14 &$\pm$ 8.71e-17 & 11.99 & a \\
Johnson:U &\textcolor{white}{00}3~971.0& 5.61e-14 &$\pm$ 2.80e-15&12.04& b  \\
Johnson:B &\textcolor{white}{00}4~481.0 &4.08e-14&$\pm$ 2.04e-15&12.39& b \\
XMM-OT:B &\textcolor{white}{00}4~505.79 &2.86e-14 &$\pm$ 1.36e-17 & 12.77&a \\
Johnson:V &\textcolor{white}{00}5~423.0 &2.06e-14&$\pm$ 1.03e-15& 13.13 & b \\
XMM-OT:V &\textcolor{white}{00}5~430.53 &1.92e-14 &$\pm$ 1.02e-16 & 13.20 &a \\
POSS-II:F &\textcolor{white}{00}6~399.94 &1.35e-14&$\pm$ 5.64e-15& 13.59 &c \\
Johnson:R &\textcolor{white}{00}6~441.0 &8.97e-15 &$\pm$ 4.49e-16& 14.03 &b \\
POSS-II:i& \textcolor{white}{00}7~836.89& 8.69e-15 &$\pm$ 3.47e-15& 14.07&c \\
Johnson:I &\textcolor{white}{00}8~071.0 &5.85e-15 &$\pm$ 2.92e-16& 14.50 &b\\
VISTA:Y &\textcolor{white}{0}10~183.86& 2.38e-15 &$\pm$ 2.89e-18& 15.48&d \\
2MASS:J &\textcolor{white}{0}12~390.17& 1.74e-15 &$\pm$ 3.12e-17& 15.81 & f \\
VISTA:J &\textcolor{white}{0}12~463.83& 1.81e-15 &$\pm$ 1.93e-18& 15.77 & d \\
Johnson:J &\textcolor{white}{0}12~500.21& 1.75e-15 &$\pm$ 3.45e-17& 15.81 & e \\
Johnson:H &\textcolor{white}{0}16~300.16& 6.79e-16 &$\pm$ 1.58e-17& 16.83 & e \\
2MASS:H &\textcolor{white}{0}16~494.77& 6.70e-16&$\pm$ 1.65e-17& 16.85 & f \\
VISTA:Ks &\textcolor{white}{0}21~337.54& 2.16e-16&$\pm$ 6.58e-19& 18.08 & d \\
2MASS:Ks& \textcolor{white}{0}21~637.85& 2.88e-16 &$\pm$ 7.68e-18& 17.77 & f \\
Johnson:K& \textcolor{white}{0}21~900.25& 2.72e-16&$\pm$ 8.75e-18& 17.83 & e \\
AKARI:N3 &\textcolor{white}{0}31~899.94& 1.06e-16&$\pm$ 3.54e-18& 18.85 & g \\
WISE:W1& \textcolor{white}{0}33~500.10& 7.83e-17 &$\pm$ 1.87e-18& 19.18 & h \\
Spitzer/IRAC:3.6\textcolor{white}{b} &\textcolor{white}{0}35~499.82& 6.07e-17 &$\pm$ 2.38e-18& 19.46 & i \\
Spitzer/IRAC:4.5 &\textcolor{white}{0}44~930.23& 3.49e-17&$\pm$ 8.91e-19& 20.06 & i \\
WISE:W2 &\textcolor{white}{0}46~000.19& 2.88e-17 &$\pm$ 7.08e-19& 20.27 & h \\
Spitzer/IRAC:5.8& \textcolor{white}{0}57~309.64& 1.51e-17 &$\pm$ 5.48e-19& 20.97 & i \\
AKARI:S7& \textcolor{white}{0}71~199.46& 7.63e-18 &$\pm$ 3.55e-19& 21.71 & g \\
Spitzer/IRAC:8.0& \textcolor{white}{0}78~720.81& 6.34e-18&$\pm$ 2.90e-19& 21.91 & i \\
AKARI:S11 &104~501.00& 2.96e-18&$\pm$ 1.92e-19& 22.73 & g \\
WISE:W3 &115~598.23&4.85e-19 &$\pm$ 1.48e-19& 24.70 & h \\
WISE:W4 &220~906.68& 1.12e-18 &$\pm$ 3.81e-19& 23.79 & h \\
        \hline
     \end{tabular}
         \flushleft
    \begin{tablenotes}
      \small
      \item \textbf{Notes.} Data from: $^{\text{(a)}}$ \cite{Page_2012}; $^{\text{(b)}}$ \cite{Bonanos_2009}; $^{\text{(c)}}$ \cite{Lasker_2008}; $^{\text{(d)}}$ \cite{Cioni_2011}; $^{\text{(e)}}$ \cite{Kato_2007}; $^{\text{(f)}}$ \cite{Zacharias_2013}; $^{\text{(g)}}$ \cite{Kato_2012}; $^{\text{(h)}}$ \cite{Cutri_2012}; $^{\text{(i)}}$ \cite{Meixner_2006}
    \end{tablenotes}
    \label{table_data_vizier}
\end{table*}

\newpage

\section{Radial velocities of the components}


\begin{table*}[]
    \centering
    \caption{Radial velocities of the four components at each epoch.}
    \begin{tabu}{c|cccc}
         \hline
         \hline
        \multirow{2}{*}{MJD$^\text{a}$} & \multicolumn{4}{c}{RV [km/s]}\\
         & WN2.5-3$^\text{b}$ & O4~V & O6.5~V & B1~V \\
         \hline
        58394.29 & $15\pm6$ & $117\pm37$ & $688\pm16$ & $263\pm10$ \\
        58411.23 & $17\pm3$ & $107\pm34$ & $660\pm96$ & $252\pm7$ \\
        58413.24 & $20\pm17$ & / & / & $283\pm10$ \\
        58437.31 & $14\pm4$ & / & / & $288\pm12$ \\
        58438.22 & $19\pm7$ & $391\pm34$ & $31\pm24$ & $270\pm5$ \\
        58444.10 & $17\pm6$ & $156\pm20$ & $577\pm36$ & $251\pm19$ \\
        58446.09 & $33\pm7$ & $475\pm31$ & $-92\pm7$ & $-235\pm15$ \\
        58446.25 & $24\pm6$ & $472\pm44$ & $-117\pm20$ & $262.8\pm0.4$ \\
        58448.28 & $31\pm5$ & $205\pm12$ & $513\pm18$ & $290\pm15$ \\
        58457.29 & $20\pm4$ & $423\pm54$ & $-3\pm14$ & $260\pm4$ \\
         \hline
    \end{tabu}
         \flushleft
       \begin{tablenotes}
      \small
      \item \textbf{Notes.} $^{\text{(a)}}$ Mid-exposure; $^{\text{(b)}}$ Relative to the coadded frame
    \end{tablenotes}
    \label{table_RVs}
\end{table*}
\end{appendix}

\end{document}